# A techno-economic model for avoiding conflicts of interest between owners of offshore wind farms and maintenance suppliers

Alberto Pliego Marugán [2], Fausto Pedro García Márquez [1], Jesús María Pinar Pérez [2]

[1]Ingenium Research Group, Universidad Castilla-La Mancha, 13071 Ciudad Real, Spain.

[2] Department of quantitative methods, CUNEF Universidad, CUNEF-Ingenium, Madrid, Spain.

alberto.pliego@cunef.edu, faustopedro.garcia@uclm.es, jesusmaria.pinar@cunef.edu

**Abstract**

Currently, wind energy is one of the most important sources of renewable energy. Offshore locations for wind turbines are increasingly exploited because of their numerous advantages. However, offshore wind farms require high investment in maintenance service. Due to its complexity and special requirements, maintenance service is usually outsourced by wind farm owners. In this paper, we propose a novel approach to determine, quantify, and reduce the possible conflicts of interest between owners and maintenance suppliers. We created a complete techno-economic model to address this problem from an impartial point of view. An iterative process was developed to obtain statistical results that can help stakeholders negotiate the terms of the contract, in which the availability of the wind farm is the reference parameter by which to determine penalisations and incentives. Moreover, a multi-objective programming problem was addressed that maximises the profits of both parties without losing the alignment of their interests. The main scientific contribution of this paper is the maintenance analysis of offshore wind farms from two perspectives: that of the owner and the maintenance supplier. This analysis evaluates the conflicts of interest of both parties. In addition, we demonstrate that proper adjustment of some parameters, such as penalisation, incentives, and resources, and adequate control of availability can help reduce this conflict of interests.

The work reported herein was supported financially by the Ministerio de Ciencia e Innovación (Spain) and the European Regional Development Fund, under Research Grant WindSound project (Ref.: PID2021-125278OB-I00).

# 1 Introduction

Over the past several decades, the wind energy industry has shown significant growth. For instance, in the European Union, more than half of new renewable installations are expected to be wind turbines (WTs), with wind being the largest resource of renewable energy in 2020, surpassing hydro power [1]. Despite the COVID-19 pandemic, 2020 was a record year for the wind energy industry [2], with 93 GW of new installations. According to the European Wind Energy Association (EWEA), wind energy is expected to meet 50% of Europe's electricity demand by 2050 [3].

Today, offshore wind energy has capacity of more than 35 GW, which represents more than 4.8% of the cumulative global wind capacity [2]. Locating wind farms offshore has multiple advantages [4]; however, the main drawback of offshore wind farms (OWF), apart from the higher initial investment, is the complexity and high cost of operation and maintenance (O&M) tasks, which can account for up to 30% of total OWF project lifecycle expenditure [5]. Moreover, these costs can account for up to 25% of the levelized cost of energy produced by an OWF. Even with good weather conditions, O&M tasks are more expensive for OWF than for onshore facilities. Onshore WTs can reach reliability values from 95% to 97%, whereas OWF have significantly lower reliability [6]. Hence, the maintenance of OWF is a current challenge and is widely researched.

In general, different maintenance strategies for OWF can be found in the literature. For example, Zhu et al. [7] proposed a dynamic programming-based maintenance model, and Sarker et al. [8] proposed a multi-level opportunistic preventive strategy. Schrotenboer et al. [9] studied the stochastic maintenance fleet transportation problem for OWFs, and Koukoura et al. [10] presented an stochastic approach for determining main key performance indicators (KPIs) to reduce the impact of condition monitoring systems on reliability. For further information, Ren et al. [11] presents a review of main strategies, methods, and other issues of O&M of OWFs. All of these maintenance strategies are tailored to improve the availability of OWFs and reduce maintenance costs. The availability depends on three main variables: system reliability [12], system maintainability [13], and the efficiency of logistics [14]. Pliego et al. [15] presented a complete fault tree–based model to demonstrate the effect of resource allocation on the availability of OWF.

Maintenance activities for OWF require important investments in new infrastructure, training technicians, and obtaining adequate equipment and tools [16]. For this reason, there is a growing trend to outsource the maintenance of OWF to specialised service companies [17].

The wind power industry initially used so-called material-based contracts, where maintenance, repair, and overhaul activities were performed by the original equipment manufacturer [18]. Under this agreement, the owner would pay the manufacturer an agreed amount when an activity was carried out. However, this approach was prone to conflicts of interest as manufacturers did not have an incentive to increase the availability of the WTs because maintenance, repair, and overhaul activities generated significant revenue for the manufacturer. To solve this conflict, a new kind of contracting, called performance-based contracting (PBC), is being adopted for maintenance service in industrial processes, whereby the owner pays the maintenance contractor for meeting desired availability thresholds [18,19]. Thus, contractors receive incentives from their clients to improve the availability of products and systems. This type of contract is used in numerous sectors, such as health care [20], construction [21], and social welfare programmes [22], among others.

Guajarado et al. [23] analysed the impact of PBC on product reliability and demonstrated that reliability is improved by 25% to 40% under PBC compared with material-based contracting. The characteristics of PBC have been widely studied in several industries. Straub et al. [24] discussed the main variables to consider when selecting performance-based maintenance partnerships. They also analysed the cost savings under PBC [25] and concluded that PBC reduces both direct and indirect costs. Numerous general studies have been conducted on PBC [26-28], as well as studies that are oriented to manage specific systems such as rest areas [29] or manufacturing systems [30]. A general literature review of PBC can be found in Selviaridis et al.[31].

However, the literature on modelling maintenance outsourcing of OWF is scarce and insufficient. Offshore wind energy systems are a special case because multiple stochastic variables are involved, which complicates the calculation of costs and profits. In fact, some papers have analysed different types of agreements and contracts [32,33] of outsourcing WT maintenance, but they do not quantify differences between the interests of stakeholders. In such papers, incentivising has been demonstrated to be an effective operation. However, it is not clear how much incentive is necessary to improve the profits of both parties [34]. The approach developed in this paper allows these issues to be addressed statistically. With respect to the wind energy industry, the literature includes several studies about the pros and cons of different maintenance contracts [35]. For instance, Jin et al. [18] presented a quantitative model that considered PBC and reached an important conclusion: the contractor invests more resources when the contract period is longer. Jin et al. [36] also proposed a game-theoretical method for maximising the utilities of all stakeholders under PBC.

In this paper, we intend to optimise some terms of maintenance contracts where control of availability is an essential requirement. This paper presents the following main objectives: first, to evaluate and quantify the conflict of interest between the owner and the maintenance supplier; and second, to demonstrate that the correct setting of penalisations, incentives, and resources can avoid such a conflict of interest. The problem is addressed through a double-objective programming problem oriented to maximise profits for both parties while minimising the conflict of interest. These objectives are reached by considering the number of technicians as well as thresholds and quantities for penalisations and incentives as decision variables. This optimisation problem provides a set of efficient solutions from which the terms of a contract can be negotiated. We propose a complete simulation model for estimating the costs and profits of OWF owners and maintenance suppliers. The data used in the case studies are based on statistical information extracted from the literature and used to create estimates required by the proposed methodology.

This paper contributes to the offshore wind industry by presenting a methodology to quantify and reduce the conflict of interest between the OWF owner and the maintenance supplier through adequate negotiation of incentives and penalisations. Most of the literature about maintenance of OWFs focuses on tools and strategies that improve maintenance. For instance, tools like supervisory control and data acquisition (SCADA) systems and condition monitoring systems [37] or strategies such as window maintenance, opportunity maintenance [38], or routine maintenance [39,40] are beneficial for both parties. However, this paper focusses not only on maintenance strategies but also (and mainly) on the terms of the maintenance contracts, including penalisation and incentives, that minimise the conflict of interest between the parties while maximising their profits.

The rest of paper is structured as follows: Section 2 analyses the main terms and variables of maintenance contracts; Section 3 presents an estimation of costs and profits from the different perspectives of stakeholders and proposes a methodology to minimise the conflict of interest between them; Section 4 explains the model created for simulating the variables and generating realistic scenarios; Section 5 presents a case study for applying the proposed methodology; Section 6 shows graphically the influence of penalisations and incentives in profits of both parties; and Section 7 presents the main conclusions of this research.

## 2 Maintenance contract terms

In this paper, we analysed the possible conflicts of interest that may occur between the owner/operator of the wind farm and the maintenance contractor. To avoid this situation, the terms of the contract may be negotiated by both parties. In general, contracts under PBC are composed of a fixed fee and a variable fee. These fees are usually established according to performance during the contractual period. The contracts are usually for a period of 5 to 10 years [32]. After this period, the conditions can be renegotiated. In some cases, contract terms are renegotiated annually [32].

The fixed fee is established according to the responsibilities assumed by the contractor. The maintenance service can be composed of different tasks included in the contract. In this paper, the contractor sets the required maintenance strategies for controlling the availability of the OWF.

2.1 Availability centred model

Maintenance is the execution of different activities to ensure correct performance of the OWF. The quality of maintenance can be quantified through different performance indicators, such as availability [41], wind/energy index, capacity factor, $P_{50}$ deviation, and so on. A complete review of key performance indicators can be found in Gonzalez et al.[42]. In this paper, we developed a model to quantify maintenance performance by considering three different availabilities: wind farm availability, WT availability, and energy-based availability .

Wind farm availability is the probability that the wind farm is operating correctly in a given period [43]. This availability, which is considered the arithmetic average of the availabilities of the WT, is defined by equation (1):

$$A^{WF} = \sum_{w \in W} \sum_{t \in T} \frac{a_{wt}}{N \cdot T} \qquad (1)$$

where:

$A^{WF}$: Wind farm availability
$a_{wt}$: Probability that WT $w$ is working correctly in period $t$.
$N$: Total number of WT in the OWF.
$T$: Total time period
$W$: Set of wind turbines

WT availability is the quotient between the hours that a WT is working and the total number of hours in the period. This availability is calculated by equation (2):

$$A_w^{WT} = \frac{\sum_{t \in T} a_{wt}}{T} \qquad (2)$$

where: $A_w^{WT}$ is WT availability.

The maintenance supplier can also undertake to control energy-based availability, which is the ratio between the energy expected and the energy effectively generated. It is defined by equation (3):

$$A^G = \frac{\sum_{w \in W} G_{wt} \cdot a_{wt}}{\sum_{w \in W} G_{wt}} \qquad (3)$$

where:

$A^G$: Energy-based availability
$G_{wt}$: Electricity generated by the WT $w$ at time $t$ [MWh]

To maintain different availabilities at appropriate values, the maintenance supplier can follow different maintenance strategies and techniques [44,45]. The main types of maintenance are corrective maintenance, also called unscheduled or failure-based maintenance, which is carried out when faults or failures are detected in a WT [37]; preventive maintenance, where specific maintenance activities are regularly performed on components to prevent breakdown of the system; and predictive maintenance, where system data are acquired and processed to detect anomalies and determine actual maintenance requirements. Further details of methods and strategies for maintenance optimisation can be found in Márquez et al. and Shafiee and Sørensen [37,46].

In this paper, we aimed not to establish the best maintenance strategy but rather to quantify the difference between the O&M strategy adopted by the contractor and the interests of the owner. The total cost of maintenance activities is given by the total costs of corrective, preventive, and predictive activities. These costs are defined by equation (4):

$$C^A = \sum_{t \in T} \sum_{w \in W} \left( \sum_{j \in J_p} C_{jwt}\, \beta_{jwt} + \sum_{j \in J_c} C_{jwt}\, \beta_{jwt} + \sum_{j \in J_d} C_{jwt}\, \beta_{jwt} \right) \quad (4)$$

where:

$C^A$: Total cost of the maintenance activities [€]
$\beta_{jwt}$ = 1 if the task *j* is executed in WT *w* at time *t*, and 0 otherwise
$C_{jw}$: Cost of a maintenance task *j* on WT *w* [€]
$J_c, J_p, J_d$: Sets of corrective, preventive, and predictive activities

The costs defined in equation (4) affect both the owner and the contractor. The cost of a certain maintenance task can be divided into three different costs: transportation costs, costs based on repair time, and costs based on materials. In the proposed model, the owner of the OWF will bear the material costs, whereas the transportation and costs based on repair time will be assigned to the contractor. The allocation of these costs will be detailed in Sections 3 and 4.

## 2.2 Contractual commitment

The contractor is usually committed to maintaining availability over a certain threshold (*minimum availability thresholds*). These thresholds allow the contractor to offer a performance guarantee. If the thresholds are not met, the contractor must pay a penalty for liquidated damages. Specific thresholds and penalisations will be fixed for each type of availability. These penalisations are commonly calculated and computed every 6 months.

The penalisation associated with $A^{WF}$ is calculated as the relative difference between the incomes of the owner when the threshold is met and when it is not met. This penalisation can be calculated by equation (5):

$$\xi^{WF} = max\left\{0, \sum_{t \in T} (I^{own}) \cdot \left(\frac{R^{WF} - A^{WF}}{R^{WF}}\right)\right\} \quad (5)$$

where

$\xi^{WF}$: Penalisation for poor $A^{WF}$ [€]
$R^{WF}$: Minimum wind farm availability threshold
$I^{own}$: Incomes for the owner [€]

Equation (5) shows that if WF availability is above the minimum threshold at all times, penalisation will be null.

In addition to the minimum wind farm availability threshold, the contractor can also guarantee a *minimum availability for each WT*. In this case, the contractor will be penalised if any WT is paused for more than a certain period. This period ($R^{WT}$) will be determined as a percentage of the total period [47]. If the WT has worked for longer than $R^{WT}$, then penalisation will be null. Therefore, a specific penalisation will be counted for each WT. The total penalisation for WT availability ($\xi^{WT}$) is given by equation (6):

$$\xi^{WT} = \sum_{w \in W} max \left\{ 0, \sum_{t \in T} \frac{I^{own}}{N} \cdot \left( \frac{R^{WT} - A_w^{WT}}{R^{WT}} \right) \right\} \qquad (6)$$

where

$\xi^{WT}$: Penalisation for poor $A^{WT}$ [€]
$R^{WT}$: Minimum WT availability threshold

In this case, the contractor would be penalised if a minimum energy-based availability threshold was not met. This penalisation will be calculated as the difference between actual and expected incomes for the owner. The total amount of the penalisation is defined by equation (7):

$$\xi^G = max \left\{ 0, \sum_{t \in T} (I^{own}) \cdot \left( \frac{R^G - A^G}{R^G} \right) \right\} \qquad (7)$$

where

$\xi^G$: Penalisation for poor energy-based availability, $A^G$ [€]
$R^G$: Minimum energy-based availability threshold

From the perspective of the owner, the penalisation for liquidated damages is an income. However, the owner does not benefit from this income because penalisations are a consequence of bad maintenance performance. The total amount of the liquidated damages is capped by an upper limit ($\lambda^{LD}$). This limit is usually a proportional value of the total amount billed by the contractor in a year. Equation (8) shows the total amount of the liquidated damages ($\xi^{LD}$) perceived by the owner in period *T*.

$$\xi^{LD} = min\ (\lambda^{LD}, \xi^{WF} + \xi^G + \xi^{WT}) \qquad (8)$$

where

$\xi^{LD}$: Total amount of liquidated damages [€]
$\lambda^{LD}$: Upper limit of liquidated damages [€]

Some exceptional events can affect the liquidated damages; for example, vandalism, natural catastrophes, or failures. In these cases, the contractor will be exempt from payment of the liquidated damages.

The previous paragraphs are based on the penalisations that the contractor assumes when maintenance performances is inadequate. At the same time, good maintenance service can be rewarded through incentives; these incentives are known as "upside sharing" [32]. The upside sharing mechanism is not symmetric to performance guarantees. The performance guarantee is a proportional estimation of losses incurred during low availability. In this case, the owner shares a part of the additional gains generated in the high-availability periods. Because the owner's profits are associated with energy generation, the remuneration from upside sharing is calculated by considering only energy-based availability. This limit will be fixed as a proportional part of the total amount billed by the contractor. The total amount of this remuneration in the period *T* is defined by equation (9):

$$\xi^{US} = min \left\{ \lambda^{US}, max \left\{ 0, \sum_{t \in T} (I^{own}) \cdot \left( \frac{R^G - R^{US}}{R^{US}} \right) \right\} \right\} \qquad (9)$$

where:

$\xi^{US}$: Total amount of upside sharings [€]
$\lambda^{US}$: Upper limit of upside sharings [€]
$R^{US}$: Minimum energy-based availability for upside sharings

The penalisation and incentive terms are important for determining the incomes and costs of the stakeholders. These terms must be negotiated so that they are appropriate to the interests of both parties. Therefore, these terms will be considered in the following sections as decision variables.

## 3 The perspective of the stakeholders

In this section, the terms presented in the previous section are used to calculate costs and profits for both the owner of the OWF and the contractor. This section analyses the individual goals of each party. To facilitate comprehension of the model, Figure 1 shows the main elements that have been considered in the cost-income model proposed in this section.

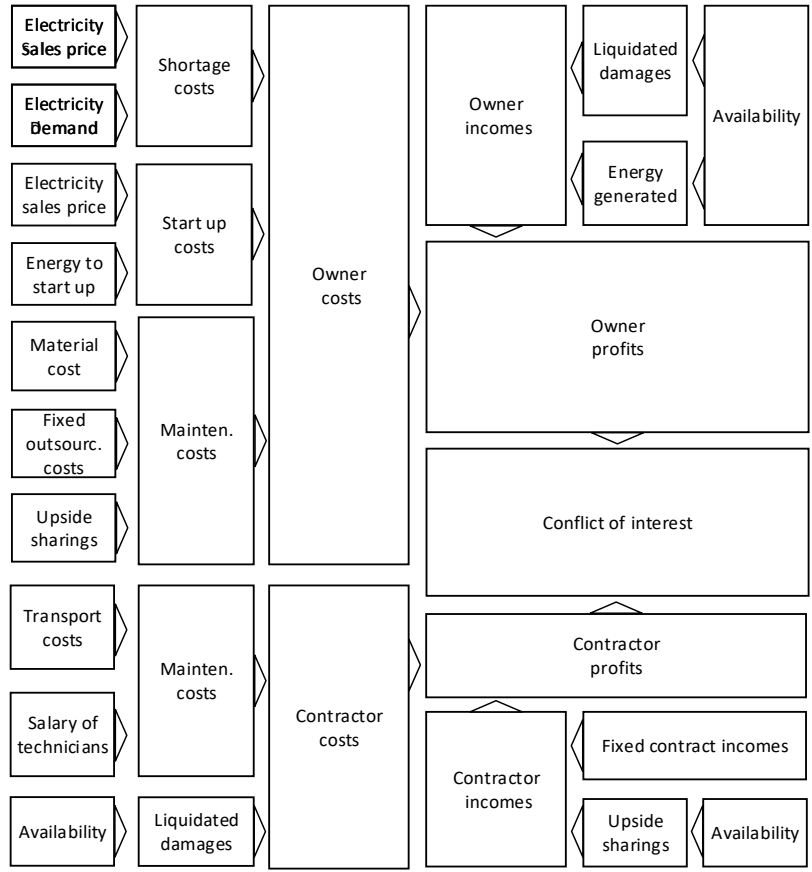

*Figure 1. Scheme of cost-income model for both parties.*

### 3.1 The perspective of the owner

The main objective of the owner is to extract maximum profits from the generated electricity. In other words, the owner wants to minimise the cost of running the wind farm and maximise the amount of energy generated and the energy sales price. According to Zhang et al. [48], the costs of running a wind farm can be expressed as the sum of three components: power shortage cost, O&M cost, and WT start-up cost. Therefore, the costs of running a wind farm can be expressed by equation (10):

$$C_t^r = C_t^s + C_t^{OM} + C_t^{up} \quad , \forall t \in T \tag{10}$$

where:

$C_t^r$: Costs of running a wind farm
$C_t^s$: Owner shortage costs at time $t$ [€]
$C_t^{OM}$: O&M costs at time $t$ [€]
$C_t^{up}$: Start-up costs at time $t$ [€]

The power shortage is the difference between the energy demand and the energy supplied by the OWF. These costs can be considered as opportunity costs caused by the unavailability of WTs.

Therefore, the power shortage costs are given by equation(11):

$$C_t^s = S_t \cdot max\left\{0, (D_t - \sum_{w \in W} G_{wt} a_{wt})\right\}, \forall t \in T \tag{11}$$

where:

$C_t^s$: Shortage costs [€]
$S_t$: Electricity sales price at time $t$ [€/MWh]
$D_t$: Energy demand at time $t$ [MWh]

The generated energy by a WT in time $t$ ($G_{wt}$) can be considered a function of predicted wind speed. The electricity sales price has been studied in different research. Interesting reviews of generated energy forecasting and electricity price forecasting can be found in [49] and [50], respectively. In this paper, we considered that all generated energy will be sold, because it is a renewable and clean energy. Therefore, the demand at each iteration time ($D_t$) is considered the maximum energy that the OWF would produce with 100% availability under the corresponding wind conditions.

WTs require energy to be activated. The start-up cost is the cost of activating a WT. In this paper, we assumed that this energy does not depend on the period when the WT has been stopped. The start-up costs are defined by equation (12):

$$C_t^{up} = \sum_{w \in W} S_t \cdot K^{UP} \cdot r_{wt} \cdot (1 - a_{w,t-1}), \forall t \in T \tag{12}$$

where:

$C_t^{up}$: Start-up costs at time $t$ [€]
$K^{UP}$: Energy required to start up a WT [MWh]
$a_{wt}$ = 1 if WT $w$ is available at time $t$, 0 otherwise.

The third major component of operating costs is O&M cost ($C_t^{OM}$). These costs can be calculated as the sum of the fixed part of the contract, the upside sharing incentives, and the material cost of each corrective maintenance task. Then, O&M costs for the owner can be estimated by equation (13):

$$C^{OM} = C^{Fix} + \xi^{US} + \sum_{t \in T} \sum_{w \in W} \sum_{j \in J} C_j^M \cdot \beta_{jwt} \tag{13}$$

where:
$C^{OM}$: O&M costs [€]
$C_j^M$: Material cost of task $j$ [€]
$C^{Fix}$: Fixed part of the contract [€]

According to equations (10)-(12), the total costs for the owner ($C^{own}$) in time period $T$ are given by equation (14):

$$C^{own} = C^{Fix} + \xi^{US} + \sum_{t \in T}\sum_{w \in W}\sum_{j \in J_c} C^M_{jwt} \cdot \beta_{jwt} + \sum_{t \in T} S_t \cdot max\left\{0, \left(D_t - \sum_{w \in W} G_{it} a_{wt}\right)\right\} \\ + \sum_{t \in T}\sum_{w \in W} S_t \cdot G_w \cdot a_{wt} \cdot (1 - a_{w,t-1})$$ (14)

The incomes for the owner ($I^{own}$) can be determined by the amount of energy sold and the sales price. The income in period *T* is given by equation (15):

$$I^{own} = \xi^{LD} + \sum_{t \in T}\sum_{w \in W} S_t \cdot G_{wt} \cdot a_{wt}$$ (15)

Therefore, the profits of the owner ($B^{own}$) in period *T* can be expressed as the difference between the total incomes and costs. The profits are defined by equation (16):

$$B^{own} = \xi^{LD} + \sum_{t \in T}\sum_{w \in W} S_t \cdot G_{wt} \cdot a_{wt} \\ - \Bigg[ C^{Fix} + \xi^{US} + \sum_{t \in T}\sum_{w \in W}\sum_{j \in J_c} C^M_{jwt} \cdot \beta_{jwt} \\ + \sum_{t \in T} S_t \cdot max\left\{0, \left(D_t - \sum_{w \in W} G_{it} a_{wt}\right)\right\} \\ + \sum_{t \in T}\sum_{w \in W} S_t \cdot G_w \cdot a_{wt} \cdot (1 - a_{w,t-1}) \Bigg]$$ (16)

In conclusion, the objective of the owner is to maximise $B^{own}$. Negotiation of the variables that contribute to increasing these profits will be analysed in the following sections.

3.2 The perspective of the contractor

The main objective of the maintenance supplier is to minimise the maintenance costs while meeting the thresholds fixed in the contract. In this case, all incomes come exclusively from the maintenance contract and the variable fees.

The responsibility of the contractor is to keep the wind farm performance meeting a minimum availability and to deliver the guarantees specified in the contract. In this paper, we establish a low threshold that will cause the cancellation of the contract. All thresholds will be set according to the performance guarantees.

In our model, the costs of materials for repairing, replacing, or overhauling the affected components or systems are borne by the owner. From the perspective of the contractor, the costs of a certain maintenance activity are considered the sum of the cost of transporting the necessary equipment and crew to the specific WT and the idling costs of the corresponding means of transport during the execution of maintenance tasks. Therefore, the cost of carrying out a certain maintenance activity from the perspective of the contractor ($C^*_{jwt}$) is given by equation (17)(*17*):

$$C^*_{jwt} = Z_w \cdot C^{Km}_u + H_j \cdot C^H_u, \forall\, w, \widehat{w} \in W, t \in T, j \in J, u \in U$$ (17)

where:
    $C^*_{jwt}$: Cost of the maintenance activity *j* at WT *w* and time *t* [€]
    $C^{Km}_u$: Cost per kilometre of means of transport *u* [€/km]
    $C^H_u$: Idling cost of means of transport *u* [€/h]
    $Z_w$: Distance from the O&M base to WT *w* [km]
    $H_j$: Duration of the task *j* [h]

In addition to the costs of performing the maintenance activity, the contractor must bear the labour cost of the technicians who make up the maintenance staff. Maintenance staff includes more than just technicians; however, we only included these types of workers since it is essential to define the maintenance strategy. The number of technicians is considered a key decision variable because it affects both parties.

Considering equations (8) and (9), the incomes and costs for the contractor are defined by equations (18) and (19), respectively:

$$I^{con} = C^{Fix} + \xi^{US} \tag{18}$$

$$C^{con} = \sum_{t \in T} \sum_{w \in W} \sum_{j \in J} (Z_w \cdot C_u^{Km} + H_j \cdot C_u^H) + \xi^{LD} + C^{Tech} \tag{19}$$

where:

$I^{con}$: Incomes for the contractor [€]
$C^{con}$: Costs for the contractor [€]
$C^{Tech}$: Labour costs of the technicians [€]

The labour cost of the technicians is calculated by equation (20):

$$C^{Tech} = Q \cdot \frac{S_q}{365} \cdot T \tag{20}$$

where:

$Q$: Total number of technicians
$S_q$: Annual salary of technicians [€]

Given the incomes and costs given in equations (18) and (19), the contractor profits ($B^{con}$) are calculated in equation (21):

$$\begin{aligned} B^{con} &= C^{Fix} + \xi^{US} - \xi^{LD} - C^{Tech} - \sum_{t \in T} \sum_{w \in W} \sum_{j \in J} C^*_{jwt} \\ &= C^{Fix} \\ &+ \sum_{t \in T} \sum_{w \in W} \left( \sum_{j \in J_p} (Z_w \cdot C_u^{Km} + C_u^H) \cdot \beta_{jwt} + \sum_{j \in J_c} (Z_w \cdot C_u^{Km} + C_u^H) \cdot \beta_{jwt} \right. \\ &\left. + \sum_{j \in J_d} (Z_w \cdot C_u^{Km} + C_u^H) \cdot \beta_{jwt} \right) + \min \left\{ \lambda^{US}, \max \left\{ 0, \sum_{t \in T} (I_t^{op}) \cdot \left( \frac{R^G - R^{US}}{R^{US}} \right) \right\} \right\} \\ &- \min (\lambda^{LD}, \xi^{WF} + \xi^G + \sum_{w \in W} \xi_w^{WT}) \end{aligned} \tag{21}$$

The contractor aims to reduce the cost of the maintenance activity while ensuring a certain level of availability. Herein, we consider that the contractor will develop the best strategy to minimise maintenance costs. Cost reduction can be determined by several operative decisions; for example, the routing and scheduling problem. This is a widely studied problem that involves optimal assignments of turbines and routes to the vessels in terms of costs [51-53].

In this paper, we aimed not to develop a new algorithm to find the optimal solution to this problem, but rather to quantify and reduce possible conflicts between the contractor and the owner with respect to their agreements. This conflict of interest is managed by following the methodology proposed in Section 3.3.

## 3.3 Quantification and reduction of the conflict of interest

Because the profits of the owner and the contractor correspond to different orders of magnitude, they are scaled so that they can be compared with each other. The scaled profit will establish the profit achieved by each party with respect to the maximum possible profit. The scaled profits range from 0 to 1 according to equation (22)

$$B_{scaled} = \frac{B_i - B_{min}}{B_{max} - B_{min}} \tag{22}$$

where:

$B_{scaled}$ : Scaled profit between 0 (no profits) and 1 (maximum profit)
$B_i$ : Profit of scenario *i* [€]
$B_{max}$ : Maximum profit in the simulation [€]
$B_{min}$ : Minimum profit in the simulation [€]

An approach to finding feasible solutions to the problem is proposed when the contributions of each decision variable to the profits are analysed graphically. These solutions can be obtaining by addressing a multi-objective programming problem.

The objective of the programming problem is to determine the conditions that minimise conflicts of interest and maximise total profits. For this purpose, the constrained double-objective programming problem is defined by equations (23) and (24), subjected to constraints from equations (25) to (30).

$$minimize \; |B_{scaled}^{cont}(Q, \lambda, R^{US}, R^{LD}) - B_{scaled}^{own}(Q, \lambda, R^{US}, R^{LD})| \tag{23}$$

$$maximize \; B^{cont}(Q, \lambda, R^{US}, R^{LD}) + B^{own}(Q, \lambda, R^{US}, R^{LD}) \tag{24}$$

*Subject to*:

$$R_{min}^{US} \leq R^{US} \leq R_{max}^{US} \tag{25}$$

$$R_{min}^{LD} \leq R^{LD} \leq R_{max}^{LD} \tag{26}$$

$$\lambda_{min} \leq \lambda \leq \lambda_{max} \tag{27}$$

$$Q_{min} \leq Q \leq Q_{max} \tag{28}$$

$$\lambda^{LD} = \lambda^{US} = \lambda \tag{29}$$

$$R^G = R^{WT} = R^{WF} = R^{LD} \tag{30}$$

Equation (23) defines the first objective to minimise the difference between the scaled benefits of both parties. This equation helps to find balanced solutions where both parties obtain the same proportion of their maximum profits. Moreover, it is important not only to ensure that both parties are working at the same level, but also that this level is providing the maximum profits. For this purpose, equation (24) represents the second objective to ensure that profits are maximised.

Equations (25), (26), and (27) provide the boundary conditions for the thresholds and limits of penalisation and incentives. Equations (28) and (29) simplify the problem by reducing the number of decision variables and the computational cost.

## 4 Simulation model

Once the costs for the stakeholders were estimated, a simulation model was developed to build a case study and analyse different scenarios. This model aimed to determine the effect of the different decision variables on the profits of both parties. In this model, only variables related to O&M were considered and, among them, those variables that affect both parties were taken as decision variables. Even though the profits are related only to maintenance, the considered variables allow terms of maintenance contracts to be fixed and negotiated. Figure 2 shows a flowchart with the basic structure of the simulation model.

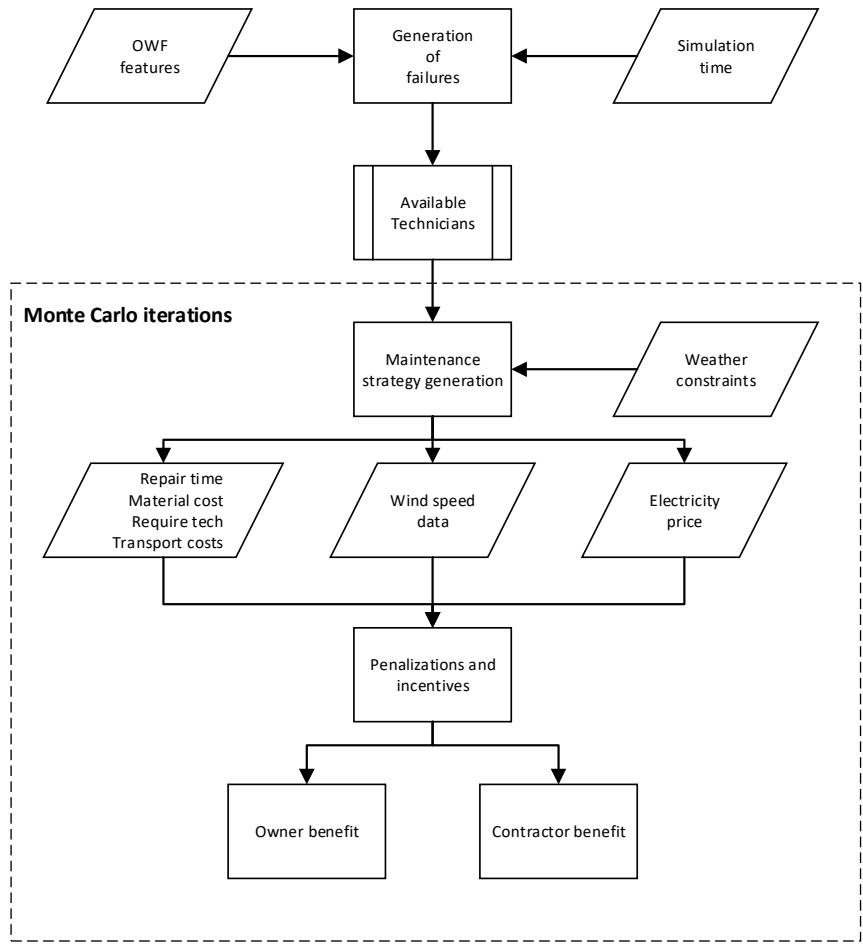

*Figure 2. Flowchart of the simulation model.*

The proposed model is based on a collection of different technical and economic data around the OWF and its stakeholders. The resulting profits and costs will not correspond to absolute results because other variables must be considered besides maintenance issues. The main characteristic of this model is that it is designed to define the dependence between the profits of both parties.

To generate a specific case study, some elements were manually inputted, such as the configuration of the wind farm and number of iterations, whereas others were stochastically generated by using a probability distribution. Table I summarises the input elements for the model.

*Table I. Summary of inputs for the simulation model.*

| Input elements of the model | | |
|---|---|---|
| **Name** | **Description** | **Type** |
| **Offshore wind farm** | Wind farm configuration, types of WT, distances, etc. | Manually pre-set |
| **Failures** | Main types of failures based on statistics, including downtimes, costs of repair, and required technicians | Stochastic |
| **Simulation time** | Total period that is considered for the simulation (discretised in days) | Manually pre-set |
| **Technicians** | Number of workers that the maintenance supplier will hire | Manually pre-set |
| **Weather conditions** | Wind speed and wave height. Some maintenance activities are constrained by these conditions. | Stochastic |
| **Electricity sales price** | The mean electricity sales price on day $t$ | Stochastic |
| **Energy demand** | The total electricity demand on day $t$ is considered the maximum energy that the OWF can generate under the wind conditions of day $t$. | Stochastic |
| **Costs of transports** | The cost of using different means of transport will be inputted to calculate the total cost of a certain activity. | Manually pre-set |
| **Penalisations and incentives** | Information regarding penalisation and incentives, such as minimum availability thresholds, upside sharing, liquidated damages, etc. | Manually pre-set |

To create a model that allows the objectives of the paper to be achieved clearly, some simplifications and assumptions have been considered. The following assumptions have been made in the proposed model.

- Only corrective maintenance is considered. In this paper, we did not intend to find the optimal maintenance strategy but to set the optimal contract terms once the maintenance strategy is fixed. Considering a more complex maintenance strategy, including preventive and predictive maintenance, would result in different maintenance costs and availabilities, although the methodology would not be affected. Therefore, this assumption will simplify the model, showing comprehensively how the costs and profits of both parties are related.
- All WTs in the wind farm are assumed to have the same power rating.
- The expected number of failures during the simulation period corresponds to the failure rates given by the literature. However, the time and the WT in which each failure occurs are randomly assigned.
- The stochasticity of the model provides different results every time the simulation is run. A Monte Carlo analysis is carried out to obtain the results.
- The wind farm is composed of a single type of WT. Therefore, all WTs have the same power.
- A different electricity sales price is considered for each day.
- The resource limitation for maintenance activities is represented by the available number of technicians, which will directly affect the costs of maintenance. All means of transports are assumed to be available when needed. The availability of vessels can be limited in real cases [9] which would impact the maintenance planning. This simplification can be relevant for maintenance scheduling optimisation, but the conflicts of interest can be analysed under this assumption.

Now that the structure of the OWF maintenance model has been explained, the following section presents a case study used to apply the methodology.

# 5  Offshore wind farm case study

This section provides the data used to create a case study through the model proposed in Section 4. These data have been collected from different references. This case study provides the information required to simulate the interaction between contractor and owner. To bring the simulation closer to reality, the values of the simulation parameters are based on statistical data extracted from several research studies. The provenance of the data will be justified in all cases to validate the results.

*Offshore Wind Farm Features:* For confidentiality reasons, the name of the OWF selected for this case study is not revealed herein. The OWF is located 17 km from the French coast, is composed of 62 Siemens SG 8.0-167 DD WTs, and occupies an area of 112 km$^2$. The distance from the computed shore to the centre of the wind farm is 29.1 km. Figure 3 shows a scheme of the OWF. Each WT has been labelled with a corresponding number. Hereafter, each WT will be referred to by these numbers.

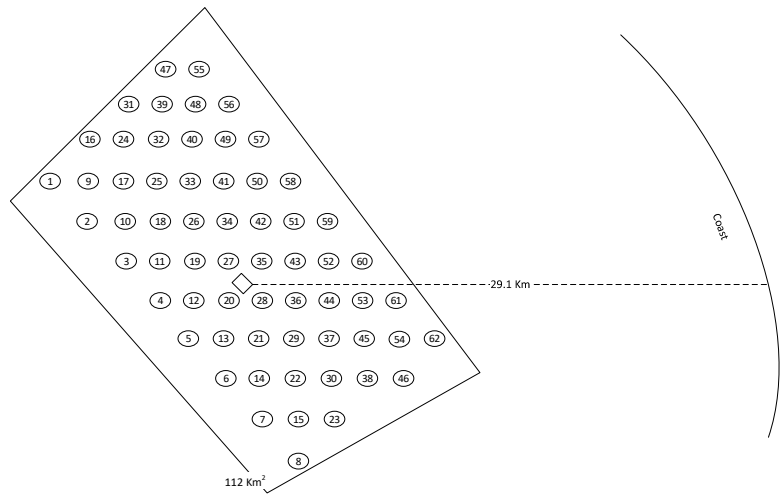

*Figure 3. Scheme of the offshore wind farm for the case study.*

*Simulation Time:* The simulation is an iterative process where the profits of both parties are calculated in all iterations. Simulation time is established by considering the terms of the contract that can be revised or renegotiated. Ferreira et al. [32] indicate that penalisations and incentives are calculated and computed every 6 months. Therefore, the analysis proposed in this paper involves a 6-month period. The iteration steps will correspond to a period of 1 day; thus, 180 iterations will be considered for the simulations.

*Generation of Failures:* To simulate the behaviour of a real OWF, a set of failures will be generated by following the statistical information in Carroll et al. [54]. Table II shows failure rates, repair time, material costs, and the number of technicians required for different failures. An identifying number has been assigned to each failure. Hereinafter, each failure will be named according to its number.

*Table II.* Data of non-scheduled activities for offshore wind turbine. Source [54]. A=major replacement (>10,000€); B=major repair (between 1000€ and 10,000€); C=minor repair (cost <1000€)

| | Failure rate (Failures/Turbine/Year) | | | Repair time (h) | | | Material costs (€) | | | Required technicians (nº) | | |
|---|---|---|---|---|---|---|---|---|---|---|---|---|
| | A | B | C | A | B | C | A | B | C | A | B | C |
| **1.Pitch/Hydraulic system** | 0.001 | 0.179 | 0.824 | 25 | 19 | 9 | 14000 | 1900 | 210 | 4 | 2.9 | 2.3 |
| **2.Other components** | 0.001 | 0.042 | 0.812 | 36 | 21 | 5 | 10000 | 2400 | 110 | 5 | 3.2 | 2 |

| | | | | | | | | | | | |
|---|---|---|---|---|---|---|---|---|---|---|---|
| 3.Generator | | 0.095 | 0.321 | 0.485 | 81 | 24 | 7 | 60000 | 3500 | 160 | 7.9 | 2.7 | 2.2 |
| 4.Gearbox | | 0.154 | 0.038 | 0.395 | 231 | 22 | 8 | 230000 | 2500 | 125 | 17.2 | 3.2 | 2.2 |
| 5.Blades | | 0.001 | 0.01 | 0.456 | 288 | 21 | 9 | 90000 | 1500 | 160 | 21 | 3.3 | 2.1 |
| 6.Grease Oil/Cooling Liq. | | 0 | 0.006 | 0.407 | 18 | 4 | 3 | 0 | 2000 | 160 | 0 | 3.2 | 2 |
| 7.Electrical components | | 0.002 | 0.016 | 0.358 | 18 | 14 | 5 | 12000 | 2000 | 100 | 3.5 | 2.9 | 2.2 |
| 8.Contactor/C. Breaker/Relay | | 0.002 | 0.054 | 0.326 | 150 | 19 | 4 | 13500 | 2300 | 260 | 8.3 | 3 | 2.2 |
| 9.Controls | | 0.001 | 0.054 | 0.355 | 12 | 14 | 5 | 13000 | 2000 | 200 | 2 | 3.1 | 2.2 |
| 10.Safety | | 0 | 0.004 | 0.373 | 0 | 7 | 2 | 0 | 2400 | 130 | 0 | 3.3 | 1.8 |
| 11.Sensors | | 0 | 0.07 | 0.247 | 0 | 6 | 8 | 0 | 2500 | 150 | 0 | 2.2 | 2.3 |
| 12.Pumps/Motors | | 0 | 0.043 | 0.278 | 0 | 10 | 4 | 0 | 2000 | 160 | 2.5 | 1.9 | 2.5 |
| 13.Hub | | 0.001 | 0.038 | 0.182 | 298 | 40 | 10 | 95000 | 1500 | 160 | 2.4 | 2.3 | 4.2 |
| 14.Heaters/Coolers | | 0 | 0.007 | 0.19 | 0 | 14 | 5 | 0 | 1300 | 485 | 0 | 3 | 2.3 |
| 15.Yaw system | | 0.001 | 0.006 | 0.162 | 49 | 20 | 5 | 12500 | 3000 | 140 | 5 | 2.6 | 2.9 |
| 16.Tower/Foundation | | 0 | 0.089 | 0.092 | 0 | 2 | 5 | 0 | 1100 | 140 | 0 | 1.4 | 2.6 |
| 17.Power Supply/Converter | | 0.005 | 0.081 | 0.076 | 57 | 14 | 7 | 13000 | 5300 | 240 | 5.9 | 2.3 | 2.2 |
| 18.Service items | | 0 | 0.001 | 0.108 | 0 | 3 | 7 | 0 | 1200 | 80 | 0 | 0 | 2.2 |
| 19.Transformer | | 0.001 | 0.003 | 0.052 | 1 | 26 | 7 | 70000 | 2300 | 95 | 1 | 3.4 | 2.5 |

The data collected in Table II have been adapted to the simulation conditions. Specifically, the data have been adapted to 62 WT and a period of 6 months. The repair time will be converted from hours into days. In this case, a repair time of 8 hours will correspond to 1 working day. This is based on a 12-hour working day of technicians, minus a 4-hour sailing out and return time [43]. Weighted averages have been used to calculate the expected number of failures, the repair time, the costs, and the number of technicians. The distribution of the failure rates for failures of types A, B, and C has been used as weights. For instance, the weights used to calculate the average cost per generator failure are 10%, 35.63%, and 53.83%, corresponding to the distribution of failures of type A, type B, and type C, respectively, in the generator. Table III shows the data of non-scheduled activities adapted to the simulation conditions.

*Table III. Data of non-scheduled activities adapted to the simulation*

| | 62 WTs & 180 days | | | | | | | |
|---|---|---|---|---|---|---|---|---|
| | Major Replace. | Major Repair | Minor Repair | Expected Number of failures | Sim. Failure rate (Failures/Turbine/day) [x 10⁻³] | Average repair time per failure (h) | Average cost per failure | Number of technicians |
| 1.Pitch/Hydraulic system | 0.031 | 5.549 | 25.544 | 31 | 2,78 | 11 | 525 | 2.41 |
| 2.Other components | 0.031 | 1.302 | 25.172 | 27 | 2,42 | 6 | 234 | 2.06 |
| 3.Generator | 2.945 | 9.951 | 15.035 | 28 | 2,51 | 21 | 7659 | 2.98 |
| 4.Gearbox | 4.774 | 1.178 | 12.245 | 18 | 1,61 | 67 | 60587 | 6.20 |
| 5.Blades | 0.031 | 0.31 | 14.136 | 14 | 1,25 | 10 | 381 | 2.17 |
| 6.Grease Oil/Cooling liq. | 0 | 0.186 | 12.617 | 13 | 1,16 | 3 | 187 | 2.02 |
| 7.Electrical components | 0.062 | 0.496 | 11.098 | 12 | 1,08 | 5 | 244 | 2.24 |
| 8.Contactor/C… | 0.062 | 1.674 | 10.106 | 12 | 1,08 | 7 | 618 | 2.35 |
| 9.Controls | 0.031 | 1.674 | 11.005 | 13 | 1,16 | 6 | 468 | 2.32 |
| 10.Safety | 0 | 0.124 | 11.563 | 12 | 1,08 | 2 | 154 | 1.82 |
| 11.Sensors | 0 | 2.17 | 7.657 | 10 | 8,96 | 8 | 669 | 2.28 |
| 12.Pumps/Motors | 0 | 1.333 | 8.618 | 10 | 8,96 | 5 | 406 | 2.42 |
| 13.Hub | 0.031 | 1.178 | 5.642 | 7 | 6,27 | 16 | 820 | 3.87 |
| 14.Heaters/Coolers | 0 | 0.217 | 5.89 | 6 | 5,38 | 5 | 514 | 2.32 |
| 15.Yaw system | 0.031 | 0.186 | 5.022 | 5 | 4,48 | 6 | 315 | 2.90 |
| 16.Tower/Foundation | 0 | 2.759 | 2.852 | 6 | 5,38 | 4 | 612 | 2.01 |
| 17.PowerSupply/Conv… | 0.155 | 2.511 | 2.356 | 5 | 4,48 | 12 | 3164 | 2.36 |
| 18.Service items | 0 | 0.031 | 3.348 | 3 | 2,69 | 7 | 90 | 2.18 |
| 19.Transformer | 0.031 | 0.093 | 1.612 | 2 | 1,79 | 8 | 1461 | 2.52 |

The data in Table III are used to generate the set of failures. Herein, major replacement, major repair, and minor repair have been merged into a single type of failure. The column "Total of failures" expresses the expected number of failures considering the sum of the three previous failure rates, a 180-day simulation period, and a total of 62 WTs. Note that this simplification can affect the real costs of maintenance operations, but it is not significant for the evaluation of conflicts of interest. Moreover, the column "Sim. Failure rates" expresses the failure rates used to stochastically generate the failures in each WT. All WTs are assumed to be operating in the period of useful life; hence, the failure rate is considered constant during the simulation period.

An example of this set is shown in Table IV, which shows the failed WT (first row), the type of failure (second row), and the day that the failure occurs (third row). A sample of the failure scenario is shown in Table IV, because the complete scenario contains all failures from Day 1 to Day 180. The number of failures is established by the failure rates given in Table III. The WTs and the day on which the failure occurs are generated randomly.

*Table IV. Example of failure scenario*

| WT      | 3 | 16 | 1  | 10 | 19 | 48 | 16 | 28 | 30 | 49 | 56 | 45 | 11 | 55 | 56 | 15 | 42 | 46 |
|---------|---|----|----|----|----|----|----|----|----|----|----|----|----|----|----|----|----|----|
| Failure | 9 | 3  | 11 | 2  | 2  | 4  | 5  | 10 | 1  | 3  | 10 | 2  | 3  | 1  | 3  | 6  | 4  | 5  |
| Day. O  | 1 | 1  | 2  | 2  | 3  | 3  | 4  | 4  | 6  | 6  | 6  | 7  | 8  | 8  | 9  | 10 | 10 | 10 |

*Number of Technicians:* According to Martin et al. [55], the number of technicians usually ranges from 0.35 to 0.75 persons per WT. Preventive and predictive maintenance are not considered in this model; therefore, fewer technicians will be required. The number of available technicians is an important variable for the generation of maintenance strategies. In this paper, the number of technicians is considered an optimisation variable. The average annual salary of the technicians ($S_q$) will be fixed at €44,000, using data from Glassdoor [56]. Salary costs will be allocated to the contractor according to equation(19). Martin et al. [55] also show statistics regarding the number of crew transfer vessels and workboats according to the size of the OFW.

*Generation of Maintenance Activities:* Once the failure scenario and the number of available technicians is known, a vector with the maintenance activities is generated. This vector, named the day of repair vector (DRV), contains the day when all maintenance activities start. This vector is randomly generated but is constrained by weather conditions, repair time, and the required technicians (see Table III). Therefore, the DRV is feasible only if enough technicians are available and the weather is adequate for all activities. Only the maintenance activities performed during the simulation period are considered for analysis. The profits of the contractor and owner will be calculated in each iteration following the DRV planning. Table V shows an example of a DRV.

*Table V. Example of the Day of Repair Vector (DRV).*

| Report of maintenance activities ||||||| 
|---|---|---|---|---|---|---|
| WT affected | Type failure | Occurrence day | Required technicians | Repair time | Available technicians | Day of repair |
| 25 | 2  | 2  | 2 | 1 | 15 | 7  |
| 11 | 2  | 3  | 2 | 1 | 15 | 13 |
| 26 | 17 | 5  | 2 | 2 | 15 | 10 |
| 6  | 17 | 6  | 2 | 2 | 15 | 14 |
| 45 | 11 | 8  | 2 | 1 | 13 | 15 |
| 11 | 1  | 9  | 2 | 2 | 13 | 10 |
| 23 | 11 | 9  | 2 | 1 | 15 | 9  |
| 20 | 2  | 10 | 2 | 1 | 15 | 12 |
| 33 | 9  | 10 | 2 | 1 | 13 | 13 |
| 17 | 2  | 11 | 2 | 1 | 11 | 13 |
| 44 | 4  | 13 | 6 | 9 | 11 | 15 |
| 24 | 9  | 14 | 2 | 1 | 5  | 15 |
| 44 | 16 | 15 | 2 | 1 | 15 | 25 |
| 45 | 3  | 15 | 3 | 3 | 9  | 18 |

*Weather Effects:* Real weather data are provided by Puertos del estado [57], which provides historical wind speed and wave height data (from 1958 to 2018) from a meteorological station placed very close to the OWF selected. According to the histograms provided by Puertos del estado [57], the main parameters are estimated to model wind speed and wave height. Wind speed is simulated by a Weibull distribution and scaled to hub height as proposed in Eurek et al. [58]. Wave height is modelled by a Gaussian distribution.

Moreover, the methodology in Taylor and Jeon [59] can be followed for a more realistic probabilistic forecasting of wave height. The weather conditions are generated by drawing a random value from the proposed probability distributions; therefore, correlations and seasonality are not considered. However, this simplification can affect the absolute values of incomes or costs, although it is not significant for the relative values between both parties and, therefore, it does not affect the objective of the paper.

Maintenance activities can only be carried out if wind and wave conditions are suitable to ensure the safety of the technicians [60]. The maximum wind speed and wave height for accessing the WT by vessel were assumed to be 10 m/s and 1.5 m, respectively, according to Besnard et al. [61]. Because most of the maintenance activities are supported by crew transfer vessel (CTV) (according to Table VI), the maintenance tasks will be postponed if the weather conditions exceed the wind speed and wave height thresholds.

*Electricity Sales Price and Demand:* Several models can be used to predict and simulate electricity prices, such as those by Lehna et al. and Mahler et al. [62,63]. However, in this paper, a price curve for electricity was generated from real data in Dahl et al. [64]. As mentioned in Section 3.1, the demand at each time is considered the maximum energy that the OWF can produce under the corresponding wind condition, assuming that all WT are available.

*Generated Energy:* Numerous models can be used to estimate the power output from wind speed data [65]. In this paper, we developed a polynomial fitting curve using the data from Table VI and the model proposed by Deshmukh [66] and by Siemens [67]. It is defined by equation (31)(31):

$$G_{wt} = \begin{cases} 0 & \Omega_t < \Omega_c, \Omega_t > \Omega_f \\ P_r \cdot \dfrac{(\Omega_t^3 - \Omega_c^3)}{(\Omega_r^3 - \Omega_c^3)} & \Omega_c \leq \Omega_t < \Omega_r \\ P_r & \Omega_r \leq \Omega_t < \Omega_f \end{cases} \quad (31)$$

Where $P_r, \Omega_c$ and $\Omega_r$ correspond to the following technical specifications given in Table VI.

*Table VI. Technical specifications of WTs [67].*

| Technical specifications of Siemens SG 8.0-167 DD | | |
|---|---|---|
| **Notation** | **Description** | **Value** |
| $P_r$ | Nominal power | 8000 kW |
| $\Omega_c$ | Cut-in speed | 3-4 m/s |
| $\Omega_r$ | Rated output speed | 12-13 m/s |
| $\Omega_f$ | Cut-out speed | 25 m/s |

*Costs of Running the Wind Farm:* Regarding the shortage costs, only those costs derived from O&M activities will be considered in this paper. The power shortage will be calculated by equation (11), and the start-up costs will be calculated using equation (12). The start-up energy ($K^{UP}$) will be set to 0.06 MW according to Zhang et al. [48].

*Transport Costs:* The transport costs are obtained from data of Tavner [43] and Carroll et al. [54]. Table VII shows the costs of different means of transports. A conversion from knots to m/s and from £ to € is carried out.

*Table VII. Cost of means of transport.*

|  | Speed (m/s) | €/h | €/km | Use Rate |
|---|---|---|---|---|
| **Crew Transfer Vessel (CTV)** | 10.20 | 81.03 | 2.21 | 96.46% |
| **Field Support Vessel (FSV)** | 6.12 | 436.23 | 19.80 | 1.18% |
| **Helicopter** | 69.87 | 888 | 3.53 | 2.36% |

The costs of each activity will be calculated according to equation (4). The costs based on the distance to the O&M base and those based on repair time are borne by the contractor. To simplify the model, the possibility of a route with multiple maintenance waypoints is not considered. Therefore, the costs based on the distance correspond to the distance from the O&M base to the specific WT and the return trip. Several maintenance tasks are allowed in the same day, but the transportation cost of each task will be calculated considering the mentioned distance. Because each maintenance task requires a different use of the vessel, the costs based on repair time will be estimated by assuming that a means of transportation is available for half of the repair time of the task activity to give an average cost. The material costs of the maintenance activities will be borne by the owner, as defined in equation (13).

*Penalisations and Incentives:* The thresholds of penalisations and incentives will be considered as optimisation variables. The penalisations and incentives are calculated using equations (8) and (9), respectively.

The number of samples from Monte Carlo iterations was determined by considering a balance between the accuracy of the solutions and the required computational time. Each sample comprised a total of 180 iterations; that is, a 6-month simulation period. The developed model was run with 2000, 10,000, and 50,000 samples, but the solution did not vary by more than 3%. Therefore, the results shown in the following section are the average outcomes from 2000 samples. The time to run each sample on the CPU was around 10 s. The simulation model was coded in MATLAB. An Intel® Core™ i5-8250 CPU 1.60 GHz processor was used in the simulation.

## 6 Results

Once the simulation model was built, different scenarios were generated to evaluate the effect of variables on the interests of the stakeholders. The decision variables are the contractor investments, represented by the number of technicians and the constraints of penalisations/incentives.

### 6.1 The contractor investment

In this paper, the contractor investment is given by the available number of technicians. We assumed that the contractor will perform maintenance activities according to the failures that occur during the simulation. Therefore, the number of technicians is used to adjust the availability of the wind farm and to obtain incentives. Higher investments will increase the capacity to solve failures during the simulation period and, consequently, availability will rise. The contractor can decide to pay higher labour costs to keep the availability meeting certain levels and receive incentives. A first approach to the effect of the number of technicians is carried out by using the statistical values given in Table VIII from Martin et al. [55] :

*Table VIII. Initial condition for simulation.*

| Nº Technicians | $\lambda^p$ | $\lambda^{US}$ | $R^{US}$ | $R^G$ | $R^{WT}$ | $R^{WF}$ |
|---|---|---|---|---|---|---|
| | 0.35 | 0.35 | 0.85 | 0.75 | 0.75 | 0.75 |

Figure 4 shows the profits of both the owner and the contractor depending on the number of technicians. The profits of the owner increase rapidly until a certain number of technicians is reached. After this point, the profits increase slightly. In this case, the owner benefits from the investments of the contractor.

The profits of the contractor involve several turning points. There is an initial drop where profits decrease because of labour costs. This trend changes when there are enough technicians to generate incentives. The contractor must find a balance between the costs of the resources, availability, and the incentives. The weather conditions and the limitation of incentives do not allow profits to be proportional to investments.

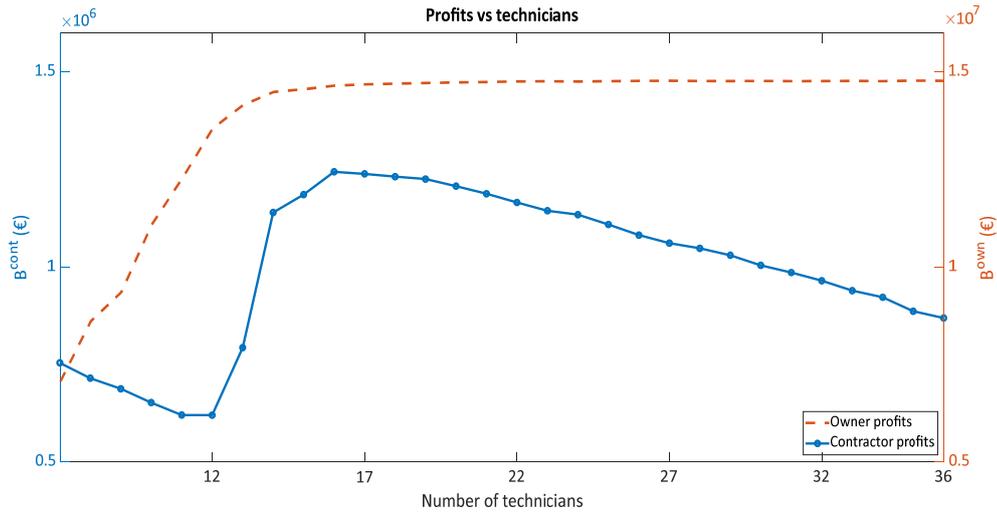

*Figure 4. Effect of the number of technicians on profits*

The maximum profit of each part is reached with different numbers of technicians. In this scenario, the maximum profits for the contractor are reached with 16 technicians, and the profits for the owner grow when the number of technicians increases. Therefore, a conflict of interest can be identified from this difference. To align the interests of both parties, the owner can induce the contractor to make a higher investment. To achieve this, the owner can improve the incentives.

6.2     The influence of penalisation and incentives

Penalisations and incentives determine the strategy that the contractor must follow to maximise their profits. These penalisations can be modified by varying the upper limits ($\lambda^p, \lambda^{US}$) and the thresholds ($R^G, R^{WT}, R^{WF}$). The variation of the thresholds means that maximum profits are reached with a different number of technicians. The total profits (sum of owner and contractor profits) do not depend on the variation in incentives. However, this variation in incentives encourages the contractor to modify the investments; therefore, total profits can vary.

Figure 5 shows the profits of the contractor versus the number of technicians and the penalisation/incentive thresholds. The red (dashed) line represents the number of technicians that maximises profits. It can be seen that a contractor would invest more resources (technicians) when the penalisation/incentive thresholds are higher. In this case, the contractor obtains a double benefit by increasing the investment: on the one hand, the minimum availability thresholds can be met to avoid penalisations; on the other hand, the upside sharing threshold could be reached, obtaining an incentive. However, if the thresholds are high (blue zone), the contractor obtains low profits and may not accept the contract conditions. If the contractor makes a poor investment, then availability would be too low (blue zone), and the owner would not accept the maintenance strategy.

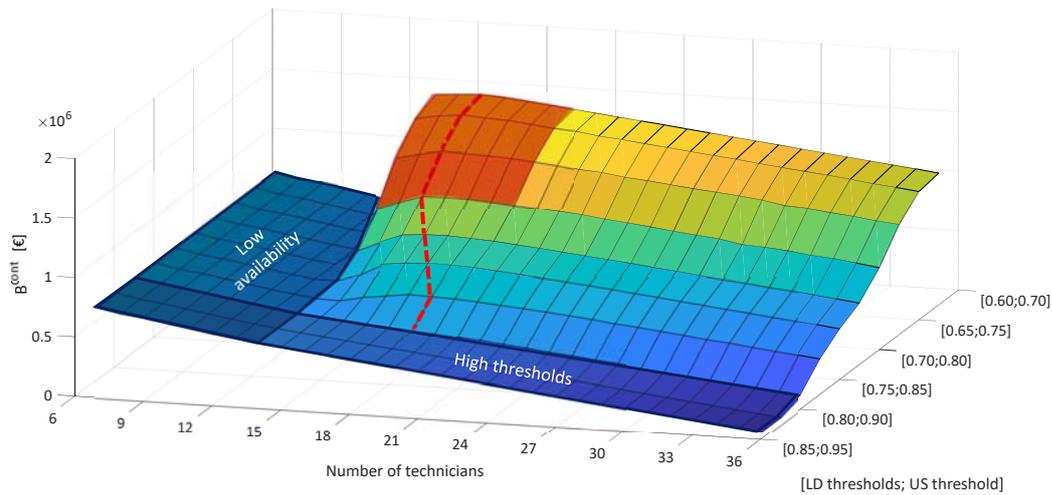

*Figure 5. Contractor profits depending on the number of technicians and the thresholds*

According to Figure 5, the contractor prefers low penalisation/incentive thresholds. However, low thresholds will cause a lower investment and low availabilities that the owner cannot tolerate. In this scenario, the contractor reaches maximum profits with 16 technicians, 0.6 liquidated damages (LD) threshold, and 0.7 upside sharing (US) threshold.

Figure 6 shows the owner depending on the number of technicians and the penalisation/incentive thresholds. The maximum profits are marked by a dashed red line. The owner is interested in maximum investment by the contractor with high penalisation/incentive thresholds.

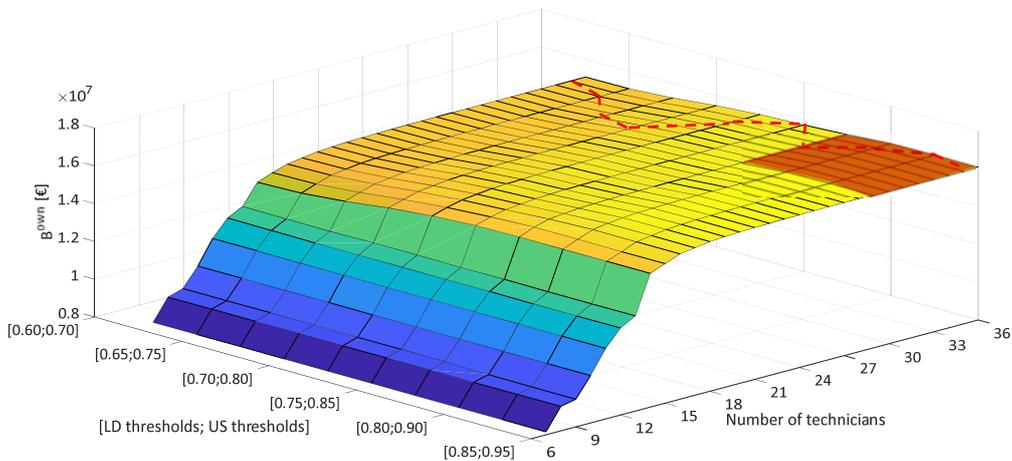

*Figure 6. Owner profits depending on the number of technicians and the thresholds*

Figure 6 shows that, from the owner's point of view, there is no clear dependence between the penalisation/incentive thresholds and the optimal number of technicians. However, maximal owner profits are reached when the number of technicians and penalisation thresholds are high (red area). In this scenario, the owner maximises their profit with 36 technicians and an LD threshold of 0.85 and US threshold of 0.95. It can be seen that the red dashed line indicating the optimum number of technicians is not a straight line. Besides the uncertainty associated with Monte Carlo, this variability indicates that, if the number of technicians is high enough, the LD and US thresholds prevail over the number of technicians; that is, the resources used by the maintenance supplier lose relevance from the perspective of the owner.

The incentives and penalisations can be also modified by varying the upper limits $\lambda^p$, $\lambda^{US}$ considered in equations (8) and (9). Figure 7 shows the contractor's profits versus the number of technicians and the upper limits.

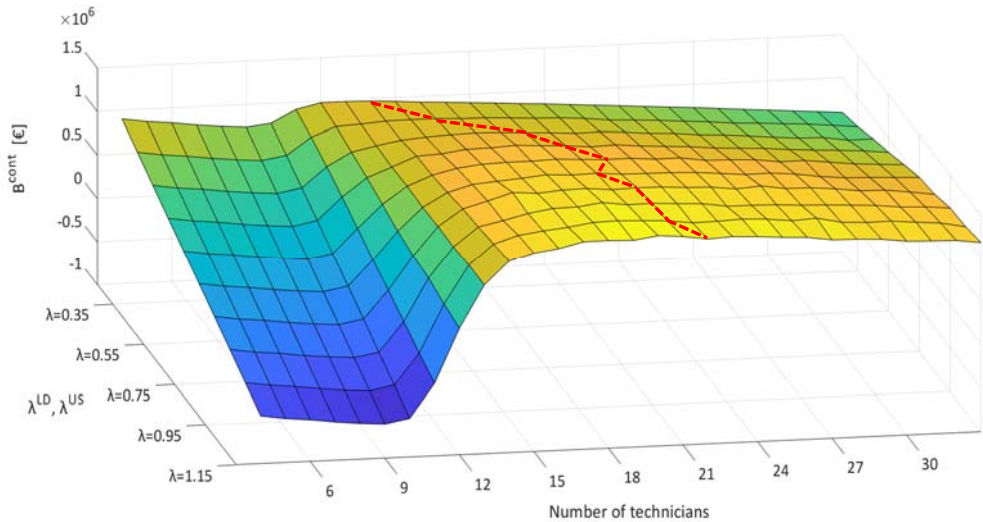

*Figure 7. Contractor profits depending on the number of technicians and the upper limits $\lambda^p$, $\lambda^{US}$.*

The increment of the upper limits means that the contractor needs more resources (technicians) to maximise profits. However, it can also mean that the contractor will have negative profits if the investment strategy does not work. Therefore, the increment of upper limits is beneficial for the contractor if they are prepared to make higher investments and take more risks.

Figure 8 shows the total profits (i.e., the sum of profits of both parties). The maximum total profit does not depend on the penalisation/incentive thresholds. However, the investment of the contractor is an important variable. From the figure, it is possible to determine the investment that the contractor should make to maximise total profits. Figure 8 shows a red band which indicates where the maximum total profits are reached.

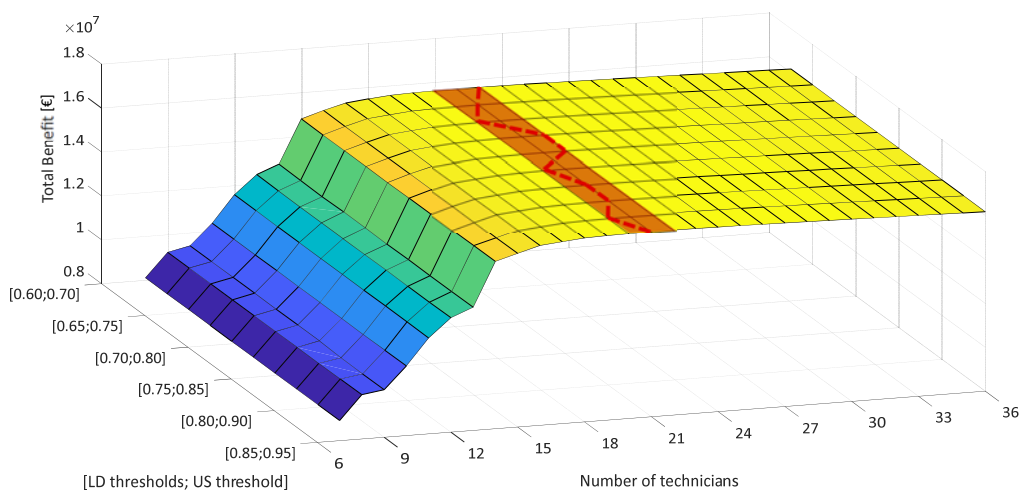

*Figure 8. Total profits versus number of technicians and minimum LD and US thresholds*

The results shown in Figure 8 correspond to a global calculation of the profitability of the OWF. Therefore, operation conditions must be adapted to maximise this total benefit. The conditions that provide the maximum total profits are not always the best because one party may have an advantage over the other.

Increasing the upper limits causes a significant response in the contractor, who will increase the number of technicians to reach these upper limits (see Figure 7); consequently, the availability of the wind farm is higher. On the other hand, when the LD and US minimum thresholds are lowered, the contractor is prompted to decrease the number of technicians (see Figure 5); subsequently, the availability of the wind farm decreases. Figure 8 shows that the maximum profitability of the OWF is reached when the contractor employs more than 20 technicians, with this number being unreachable by lowering the LD and US thresholds. We can conclude that increasing the upper limits is more profitable for the wind farm than lowering the minimum availability thresholds.

The next step is to quantify the conflict of interest and to establish equal terms for contractor and owner.

### 6.3 Multi-objective optimisation

In this section, we address the conflict of interest between the owner and the contractor by following the methodology described in Section 3.3. The objective is to determine conditions that generate a scenario in which both parties operate under the same beneficial conditions.

To determine the number of technicians that maximises the common interest, the sum of both scaled profits are considered (Figure 9). Point A corresponds to the maximum scaled profit for the contractor (16 technicians in this case study); point B indicates the maximum scaled profit for the owner (36 technicians in this case); and point C shows the maximum sum of scaled profits (20 technicians in this case study).

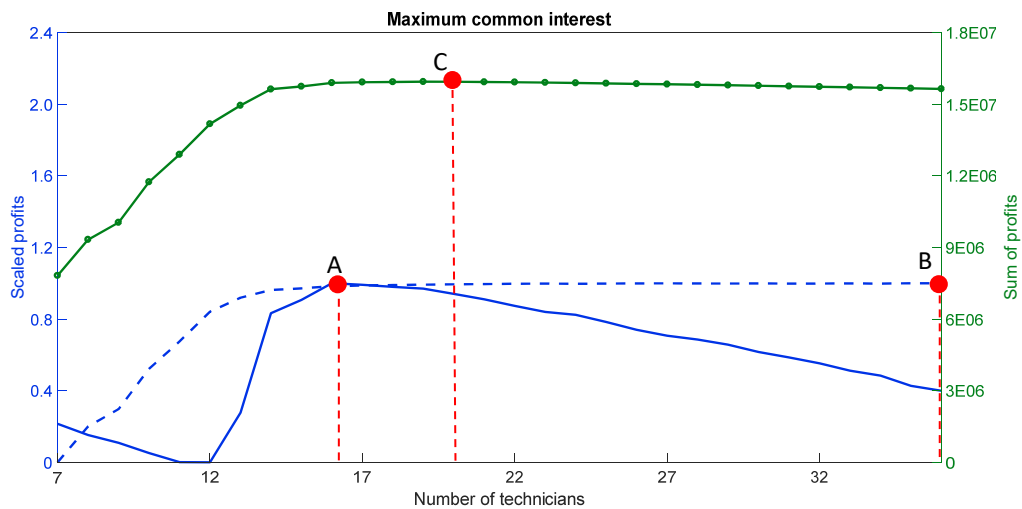

*Figure 9. Number of technicians for maximum common interest. Blue lines: Owner profit (dashed) and contractor profit scaled. Green line: sum of profits.*

The maximum total profit is reached at point C, when the sum of both profits is maximal. Under these conditions, the highest total profit in absolute terms is reached. However, this point does not coincide with the individual interests of each party (points A and B); that is, a conflict of interest exists, which can be considered as the difference between the scaled profits of each party.

Figure 10 shows the scaled profits of the contractor (red) and the owner (blue) according to the penalisation/incentive thresholds. The white dashed line shows the zones where both parties would be operating without conflict of interest (i.e., both parties are working in the same proportion of their maximum profits).

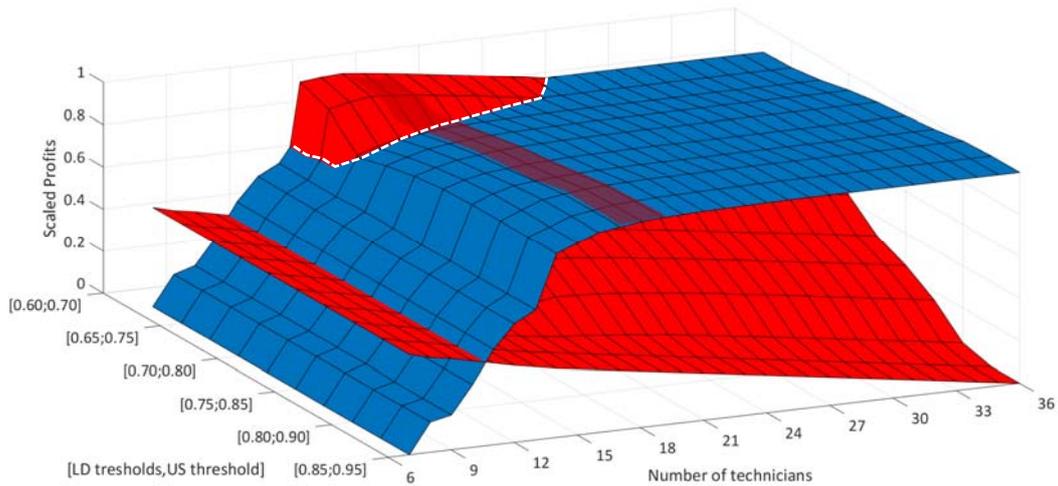

*Figure 10. Scaled profits versus Number of technicians and minimum LD, US thresholds. Blue surface: Owner's scaled profits. Red surface: Contractor's scaled profits. White (dashed) line: No conflict of interest*

Figure 11 shows the scaled profits of the contractor (red) and the owner (blue) according to the upper limits. The dashed white line shows the zone without conflicts.

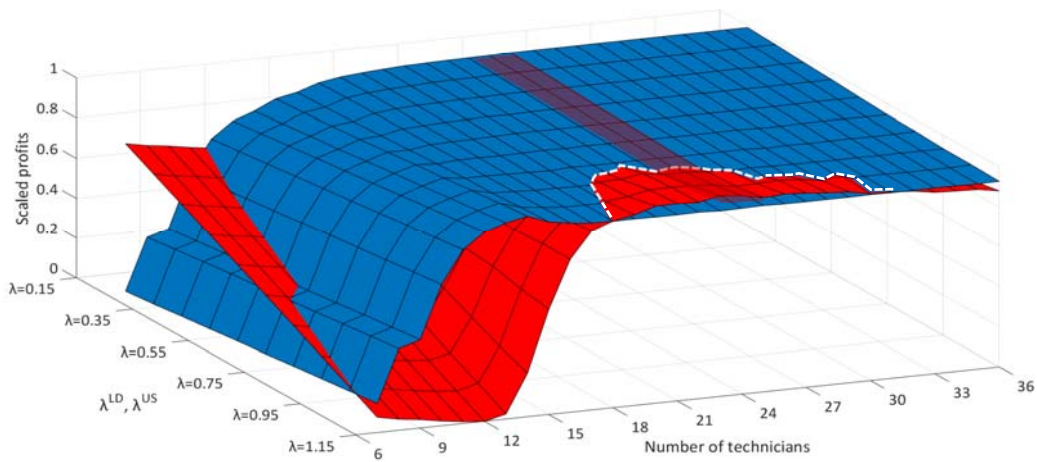

*Figure 11. Scaled profits versus Number of technicians and upper limits $\lambda^{LD}$, $\lambda^{US}$. Blue surface: Owner's scaled profits. Red surface: Contractor's scaled profits. White (dashed) line: No conflict of interest*

The dark red surfaces of Figures 9 and 10 show the zones where the total profits are maximum. Considering both objectives, maximising the total benefits and minimising the possible conflict of interests, the solutions adopted should be derived from the intersection between the dark red surface and the dashed white line.

Once the scaled profits have been obtained, a multi-objective genetic algorithm is used to solve the programming problem proposed in Section 3.3. This programming problem contains numerous stochastic variables such as failures, weather constraints, electricity sales price, energy demands, and wind speed. The problem is not based on smooth functions, and multiple local optimum points can be present. Moreover, the number of parameters is too large to be optimised by a conventional optimisation method. Multi-objective genetic algorithms have been demonstrated to be effective for problems with these characteristics; see García Márquez [68] and Zolpakar et al. [69].

The main parameter values that were used in the multi-objective genetic algorithm are summarised in Table IX.

*Table IX. Main parameters of multi-objective genetic algorithm.*

| Parameter | Description | |
|---|---|---|
| **Population size** | Number of individuals in a population | *200* |
| **Tolerance** | The algorithm stops when this tolerance is greater than 0.0001 or equal to the average relative change in the best fitness function value over the maximum stall generation. | *0.0001* |
| **Maximum stall generations** | The number of generations that are considered to determine whether the algorithm should stop according to the tolerance criterion. | *100* |
| **Max. generations** | The algorithm stops when this number of generations is reached. | *800* |
| **Mutation probability** | The probability that a mutation will be included in the next generation | *Variable* |
| **Crossover probability** | The fraction of the next generation that is created by a crossover process. The elite children are not included in this fraction. | *0.8* |

All the objective functions must be redefined as minimisation functions to facilitate the calculus. For this purpose, equation (24) is converted into equation (32):

$$minimize -B^{cont}(Q, \lambda, R^{US}, R^{LD}) - B^{own}(Q, \lambda, R^{US}, R^{LD}) \quad (32)$$

The values of the boundary conditions were assigned by using Figure 6 and Figure 8. These values are as follows:

$R^{US}_{min} = 0.50$; $R^{LD}_{min} = 0.60$; $\lambda_{min} = 0.25$; $Q_{min} = 7$;

$R^{US}_{max} = 0.85$; $R^{LD}_{max} = 0.95$; $\lambda_{max} = 1.15$; $Q_{max} = 46$;

Table X shows the result solutions of the $\Psi$. The grey columns show the objective functions evaluated with the Pareto solutions.

*Table X. Pareto optimal set.*

| | $R^{US}$ | $R^{LD}$ | $\lambda$ | Tech | Obj. 1 | Obj. 2 |
|---|---|---|---|---|---|---|
| **Solution 1** | 0.751033 | 0.793940 | 0.288468 | 23.94193 | 6.0419E-06 | 15255383.2 |
| **Solution 2** | 0.661815 | 0.718103 | 0.603721 | 27.34006 | 0.00104061 | 18577643.5 |
| **Solution 3** | 0.704413 | 0.817848 | 0.403259 | 26.49567 | 1.9468E-05 | 16300551.5 |
| **Solution 4** | 0.623042 | 0.679189 | 0.588359 | 22.42033 | 0.03627416 | 19492480.5 |
| **Solution 5** | 0.712087 | 0.703723 | 0.512941 | 18.05121 | 0.42839238 | 20039864.3 |
| **Solution 6** | 0.657295 | 0.741869 | 0.730307 | 32.40792 | 0.00062719 | 17133316.8 |
| **Solution 7** | 0.699978 | 0.781844 | 0.461220 | 27.16036 | 0.74325435 | 20238903.9 |
| **Solution 8** | 0.626146 | 0.680928 | 0.507733 | 24.40784 | 0.23265378 | 19955746.3 |
| **Solution 9** | 0.688209 | 0.632001 | 0.694395 | 22.89640 | 0.09329032 | 19540837.5 |
| **Solution 10** | 0.659123 | 0.70999009 | 0.551804 | 16.56258 | 0.01439529 | 19104818.9 |
| **Solution 11** | 0.626146 | 0.68092847 | 0.507733 | 24.40784 | 0.00406444 | 18794338.3 |

Figure 12 shows the Pareto front obtained. These solutions show the points where it is not possible to improve both objectives. In multi-objective analysis, they are called efficient solutions.

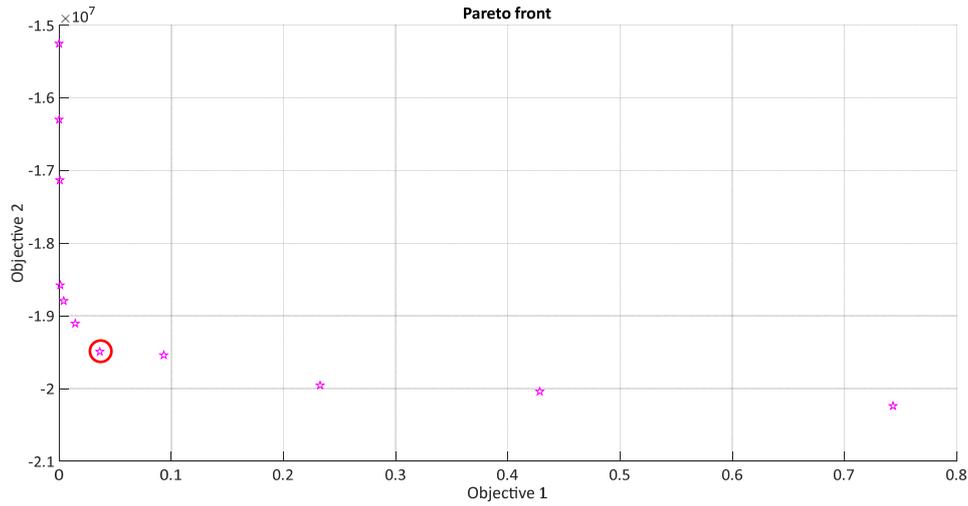

*Figure 12. Pareto front: Efficient solutions of double-objective programming problem.*

The conditions fixed in the contract should correspond to any efficient solution. There is no ideal solution that provides the best solution for both objectives. However, different criteria can be used to select a solution from among all Pareto solutions. For instance, a compromise solution that minimises the distance to the ideal point (infeasible solution where all objectives are optimal) is solution 4 (red circle). This solution is to assign the values 0.62 to $R^{US}$, 0.68 to $R^{LD}$, and 0.59 to $\lambda^{US}$ and $\lambda^{LD}$. This solution provides a high sum of profits (1.94 × $10^7$ €) and a small difference in the scaled profits of both parties (around 3%). In other words, both parties are obtaining high profits and working at similar levels of their maximum reachable profits.

6.4    Statistical uncertainty in the results

All of the presented results were obtained through Monte Carlo iterations. To analyse the error of these results, two indicators were selected: (1) profit of the contractor, and (2) total electricity generation, which is considered to calculate the profit of the owner. The error analysis was carried out by considering 2000 iterations for different numbers of technicians. Figure 13 shows the boxplot of these indicators.

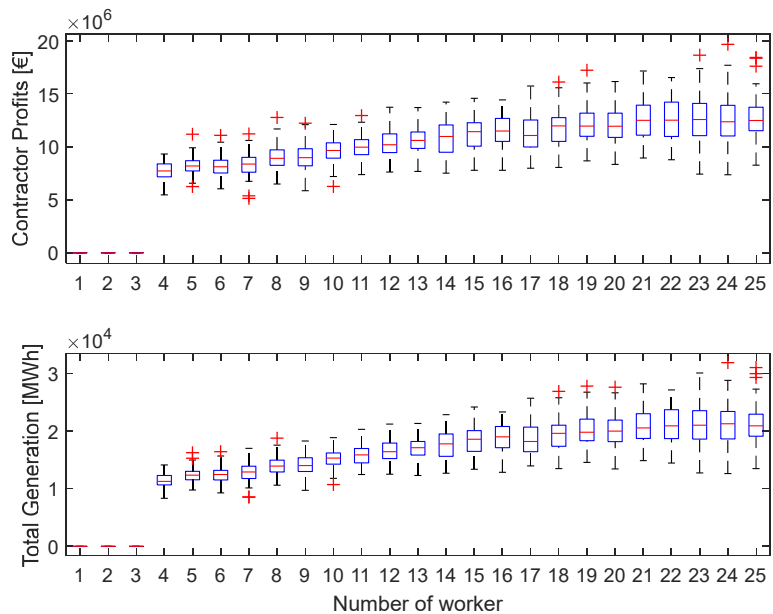

*Figure 13. Uncertainty analysis of results*

It can be seen that the variance increases slightly with the number of workers. By analysing these indicators, the following margins of error were obtained with a 95% confidence level. Depending on the number of workers, the margin of error for contractor profits varies from 1.99% to 3.5%, whereas that for total electricity generation ranges from 1.8% to 3.38%. In general, it can be assumed that the obtained results have a maximum margin of error of 3.5% with a 95% confidence level.

According to the error analysis, the methodology provides sufficient accuracy to set the correct values for the main decision variables. For example, in Figure 10 and Figure 11, the penalisations/incentives are increasing by 0.05 and 0.1, respectively, and the objective functions do not suffer abrupt changes due to these increments. Therefore, an error of 3.5% is assumed to ensure the correct operation area.

**Conclusions**

In this paper, we demonstrate the existence of a conflict of interest between the owner of a wind farm and the maintenance supplier in performance-based contracts. In addition, we show that this conflict can be attenuated by correct negotiation of incentives and penalisations in the maintenance contract. The resolution of this conflict of interest implicitly guarantees the control of availability, because availability is considered the reference parameter by which to determine which incentives or penalisations will be assigned.

The conflict of interest between both parties was quantified herein in a case study through a complete techno-economic model. The availability of the wind farm is the cornerstone of this model because it is the indicator that determines when incentives or penalisations are activated. For a clearer presentation of the methodology, we have included the following simplifications in the simulation model: only corrective maintenance is considered, a single type of WT in the wind farm, a single type of maintenance task for each component and single-waypoint routes. These simplifications might affect to the absolute profits or the availability, but they do not impede our objective of evaluating and reducing the conflict of interests. From our analysis of the statistical results, the following conclusions have been extracted.

- The first investments of the contractor, represented by the number of technicians, increase the profits of the owner rapidly. After a certain point, the profits increase slightly because of owner benefits from the investments of the contractor.
- The profits of the contractor show several turning points. First there is a gap where the profits decrease because of labour costs. This trend changes when the investment is sufficient to generate incentives. The contractor must balance costs of the resources, availabilities, and incentives to obtain maximum profits.
- The owner can promote investment by the contractor by increasing incentives. In this case, we demonstrated that increasing the upper limits of incentives is more effective than pushing down the availability thresholds.
- Even though high penalisation thresholds are usually accompanied by high incentive thresholds, the contractor would invest more resources when the penalisation thresholds are high. In this case, the contractor benefits twice by increasing investment: (1) penalisations are avoided, and (2) incentives can be achieved.
- All solutions provided by the proposed methodology imply that wind farm availability and energy-based availability are maintained above the minimum threshold for upside sharings. If the contractor does not use sufficient resources and if availability is not maintained above the threshold, then the contractor will not receive incentives. We conclude that an adequate threshold setting in the contract terms leads to improved availability.

In addition to analysis of individual interest, a multi-objective programming problem was used to find solutions that provide the best conditions for both parties. The solution of this problem consists of determining the penalisations, incentives, and resources that should be assigned to maximise the profits of both parts and align their interests. It is important to note that there is no ideal solution that simultaneously maximises global profits and minimises the conflict of interest because they are conflicting objectives. In this paper, we propose adoption of a compromise solution; that is, the efficient solution that minimises the distance with the ideal point. This solution ensures that both objectives are being considered and, therefore, the global profits are sufficiently high while the conflict of interest is reduced.

The proposed solution implicitly guarantees the improvement of availability; otherwise, the profits of the owner will be adversely affected. In future studies, the proposed methodology could be applied to more sophisticated and complex simulation models of specific OWFs.

## Nomenclature

The following notations are used in this paper.

**Sets:**

| | |
|---|---|
| **J** | Maintenance activities |
| **J$_c$** | Corrective maintenance tasks ($J_c \subset \mathbf{J}$) |
| **J$_p$** | Preventive maintenance tasks ($J_p \subset \mathbf{J}$) |
| **J$_d$** | Predictive maintenance tasks ($J_d \subset \mathbf{J}$) |
| **T** | Time periods with *t* as index |
| **U** | Means of transports with *u* as index |
| **W** | WTs at the wind farm with *w* as index |
| **Wc** | WTs that require maintenance ($W_C \subset \mathbf{W}$) |

**Parameters:**

| | |
|---|---|
| $A^{WF}$ | Wind farm availability |
| $A^{WT}_w$ | Availability of WT *w* |
| $A^G$ | Energy-based availability |
| $B^{own}$ | Total profits of the owner in period *T* [€] |
| $B^{cont}$ | Total profits of the contractor in period *T* [€] |
| $B_{scaled}$ | Scaled profit between 0 (no profits) and 1 (maximum possible profit) |
| $C^A$ | Total cost of the maintenance activities [€] |
| $C^{own}$ | Total costs for the owner in period *T* [€] |
| $C^{cont}$ | Total costs for the contractor in period *T* [€] |
| $C^{op}_t$ | Operating costs of the wind farm at time *t* [€] |
| $C^{up}_t$ | Start-up costs at time *t* [€] |
| $C^{OM}_t$ | Operation and maintenance costs at time *t* [€] |
| $C_{jwt}$ | Total cost of the task *j* in WT *w* at time *t* [€] |
| $C^*_{jwt}$ | Cost for the contractor of the maintenance activity *j* [€] |
| $C^M_j$ | Material cost of the task *j* [€] |
| $C^r_t$ | Costs of running the wind farm at time *t* [€] |
| $C^s_t$ | Power shortage costs at time *t* [€] |
| $C^{Fix}$ | Total amount of the fixed fee of the contract [€] |
| $C^H_u$ | Hourly cost of means of transport *u* [€/h] |
| $C^{Km}_u$ | Cost per kilometre of means of transport *u* [€/km] |
| $C^H_u$ | Idling cost of the means of transport *u* [€/h] |
| $D_t$ | Energy demand at time *t* [MWh] |
| $H_j$ | Duration of the task *j* [h] |
| $G_{wt}$ | Energy generated by WT *w* at time *t* [MWh] |
| $I^{own}$ | Incomes of the owner in period $T$ [€] |
| $I^{cont}$ | Incomes of the contractor in period $T$ [€] |
| $K^{UP}$ | Energy required to start up a WT [MWh] |
| $N$ | Total number of WTs |
| $R^{WF}$ | Minimum availability threshold of the wind farm |
| $R^{WT}$ | Minimum availability threshold of the WTs |
| $R^{G}$ | Minimum energy-based availability threshold of the wind farm |
| $R^{US}$ | Minimum energy-based availability for upside sharing remuneration |
| $St$ | Electricity sales price at time $t$ [€/MWh] |
| $Sq$ | Annual salary of technician [€/year] |
| $Z_w$ | Distance from the port to WT $w$ [Km] |
| $a_{wt}$ | = 1 if the WT $w$ is operating at $t$ |
| | = 0 if the WT $w$ is stopped at $t$ |
| $\beta_{jwt}$ | = 1 if the task $j$ is being executed in WT $w$ at $t$ |
| | = 0 if the task $j$ is not being executed in WT $w$ at $t$ |
| $\Omega_t$ | Wind speed at $t$ [m/s] |
| $\Psi$ | Pareto optimal set |
| $Q$ | Number of technicians |
| $\lambda^{LD}$ | Maximum compensation for liquidated damages [€] |
| $\lambda^{US}$ | Maximum upside sharing incentive [€] |
| $\xi^{WF}$ | Penalisation for availability of the wind farm [€] |
| $\xi^{WT}$ | Penalisation for availability of the WTs [€] |
| $\xi^{G}$ | Penalisation for energy-based availability of the wind farm [€] |
| $\xi^{US}$ | Incentive for upside sharings [€] |

## Acknowledgements

The work reported herein was supported financially by the Dirección General de Universidades, Investigación e Innovación of Castilla-La Mancha, under Research Grant ProSeaWind project (Ref.: SBPLY/19/180501/000102).

## References

1. Márquez, F.P.G.; Karyotakis, A.; Papaelias, M. *Renewable energies: Business outlook 2050*; Springer: 2018.
2. Council, G.W.E. GWEC| GLOBAL WIND REPORT 2021. **2021**.
3. Brindley, I.K.D.F.G. *European Wind Energy Association (EWEA). Wind energy in Europe in 2018. Trends and statitics.* ; 2019.
4. Esteban, M.D.; Diez, J.J.; López, J.S.; Negro, V. Why offshore wind energy? *Renewable Energy* **2011**, *36*, 444-450.
5. Nielsen, J.J.; Sørensen, J.D. On risk-based operation and maintenance of offshore wind turbine components. *Reliability engineering & system safety* **2011**, *96*, 218-229.
6. Cevasco, D.; Koukoura, S.; Kolios, A. Reliability, availability, maintainability data review for the identification of trends in offshore wind energy applications. *Renewable and Sustainable Energy Reviews* **2021**, *136*, 110414.
7. Zhu, W.; Castanier, B.; Bettayeb, B. A dynamic programming-based maintenance model of offshore wind turbine considering logistic delay and weather condition. *Reliability Engineering & System Safety* **2019**, *190*, 106512.
8. Sarker, B.R.; Faiz, T.I. Minimizing maintenance cost for offshore wind turbines following multi-level opportunistic preventive strategy. *Renewable energy* **2016**, *85*, 104-113.
9. Schrotenboer, A.H.; Ursavas, E.; Vis, I.F. Mixed Integer Programming models for planning maintenance at offshore wind farms under uncertainty. *Transportation Research Part C: Emerging Technologies* **2020**, *112*, 180-202.
10. Koukoura, S.; Scheu, M.N.; Kolios, A. Influence of extended potential-to-functional failure intervals through condition monitoring systems on offshore wind turbine availability. *Reliability Engineering & System Safety* **2021**, *208*, 107404.
11. Ren, Z.; Verma, A.S.; Li, Y.; Teuwen, J.J.; Jiang, Z. Offshore wind turbine operations and maintenance: A state-of-the-art review. *Renewable and Sustainable Energy Reviews* **2021**, *144*, 110886.
12. Marugán, A.P.; Chacón, A.M.P.; Márquez, F.P.G. Reliability analysis of detecting false alarms that employ neural networks: A real case study on wind turbines. *Reliability Engineering & System Safety* **2019**, *191*, 106574.
13. de la Hermosa González, R.R.; Márquez, F.P.G.; Dimlaye, V. Maintenance management of wind turbines structures via mfcs and wavelet transforms. *Renewable and Sustainable Energy Reviews* **2015**, *48*, 472-482.
14. Sobral, J.; Kang, J.; Soares, C.G. Weighting the influencing factors on offshore wind farms availability. *Advances in Renewable Energies Offshore; Guedes Soares, C., Ed.; Taylor & Francis: London, UK* **2019**, 853-863.

15. Pliego Marugán, A.; García Márquez, F.P.; Pinar Perez, J.M. Optimal maintenance management of offshore wind farms. *Energies* **2016**, *9*, 46.
16. Tautz-Weinert, J.; Yürüşen, N.Y.; Melero, J.J.; Watson, S.J. Sensitivity study of a wind farm maintenance decision-A performance and revenue analysis. *Renewable energy* **2019**, *132*, 93-105.
17. Shafiee, M. Maintenance logistics organization for offshore wind energy: Current progress and future perspectives. *Renewable Energy* **2015**, *77*, 182-193.
18. Jin, T.; Ding, Y.; Guo, H.; Nalajala, N. Managing wind turbine reliability and maintenance via performance-based contract. In Proceedings of the Power and Energy Society General Meeting, 2012 IEEE, 2012; pp. 1-6.
19. Töppel, J.; Tränkler, T. Modeling energy efficiency insurances and energy performance contracts for a quantitative comparison of risk mitigation potential. *Energy Economics* **2019**, *80*, 842-859.
20. Loevinsohn, B. *Performance-based contracting for health services in developing countries: a toolkit*; World Bank Publications: 2008.
21. Gruneberg, S.; Hughes, W.; Ancell, D. Risk under performance-based contracting in the UK construction sector. *Construction Management and Economics* **2007**, *25*, 691-699.
22. Heinrich, C.J.; Choi, Y. Performance-based contracting in social welfare programs. *The American Review of Public Administration* **2007**, *37*, 409-435.
23. Guajardo, J.A.; Cohen, M.A.; Kim, S.-H.; Netessine, S. Impact of performance-based contracting on product reliability: An empirical analysis. *Management Science* **2012**, *58*, 961-979.
24. Straub, A.; van Mossel, H.J. Contractor selection for performance-based maintenance partnerships. *International Journal of Strategic Property Management* **2007**, *11*, 65-76.
25. Straub, A. Cost savings from performance-based maintenance contracting. *International Journal of Strategic Property Management* **2009**, *13*, 205-217.
26. Xiang, Y.; Zhu, Z.; Coit, D.W.; Feng, Q. Condition-based maintenance under performance-based contracting. *Computers & Industrial Engineering* **2017**, *111*, 391-402.
27. Fuller, J.; Brown, C.J.; Crowley, R. Performance-Based Maintenance Contracting in Florida: Evaluation by Surveys, Statistics, and Content Analysis. *Journal of Construction Engineering and Management* **2017**, *144*, 05017021.
28. Jing, H.; Tang, L.C. A Risk-based Approach to Managing Performance-based Maintenance Contracts. *Quality and Reliability Engineering International* **2017**, *33*, 853-865.
29. Powers, M. Performance-Based Contracting for Rest Area Maintenance. **2017**.
30. Patra, P.; Kumar, U.D. Analysing performance-based contract for manufacturing systems using absorbing state Markov chain. *International Journal of Advanced Operations Management* **2017**, *9*, 37-56.
31. Selviaridis, K.; Wynstra, F. Performance-based contracting: a literature review and future research directions. *International Journal of Production Research* **2015**, *53*, 3505-3540.
32. Ferreira, R.S.; Feinstein, C.D.; Barroso, L.A. Operation and Maintenance Contracts for Wind Turbines. In *Use, Operation and Maintenance of Renewable Energy Systems*; Springer: 2014; pp. 145-181.
33. Wang, J.; Zhao, X.; Guo, X. Optimizing wind turbine's maintenance policies under performance-based contract. *Renewable energy* **2019**, *135*, 626-634.
34. Sandborn, P.; Kashani-Pour, A.; Goudarzi, N.; Lei, X. Outcome-based contracts–towards concurrently designing products and contracts. *Procedia CIRP* **2017**, *59*, 8-13.
35. Parlane, S.; Ryan, L. Optimal contracts for renewable electricity. *Energy Economics* **2020**, *91*, 104877.
36. Jin, T.; Tian, Z.; Xie, M. A game-theoretical approach for optimizing maintenance, spares and service capacity in performance contracting. *International Journal of Production Economics* **2015**, *161*, 31-43.
37. Márquez, F.P.G.; Tobias, A.M.; Pérez, J.M.P.; Papaelias, M. Condition monitoring of wind turbines: Techniques and methods. *Renewable energy* **2012**, *46*, 169-178.
38. Zhang, C.; Gao, W.; Yang, T.; Guo, S. Opportunistic maintenance strategy for wind turbines considering weather conditions and spare parts inventory management. *Renewable energy* **2019**, *133*, 703-711.
39. Yürüşen, N.Y.; Rowley, P.N.; Watson, S.J.; Melero, J.J. Automated wind turbine maintenance scheduling. *Reliability Engineering & System Safety* **2020**, *200*, 106965.
40. Cavalcante, C.A.; Lopes, R.S.; Scarf, P.A. Inspection and replacement policy with a fixed periodic schedule. *Reliability Engineering & System Safety* **2021**, *208*, 107402.
41. Peco Chacón, A.M.; Segovia Ramírez, I.; García Márquez, F.P. State of the Art of Artificial Intelligence Applied for False Alarms in Wind Turbines. *Archives of Computational Methods in Engineering* **2021**, 1-25.
42. Gonzalez, E.; Nanos, E.M.; Seyr, H.; Valldecabres, L.; Yürüşen, N.Y.; Smolka, U.; Muskulus, M.; Melero, J.J. Key performance indicators for wind farm operation and maintenance. *Energy Procedia* **2017**, *137*, 559-570.
43. Tavner, P. Offshore wind turbines: reliability. *Availability and Maintenance, The Institution of Engineering and Technology, London, UK* **2012**.
44. García Márquez, F.P.; Bernalte Sanchez, P.J.; Segovia Ramírez, I. Acoustic inspection system with unmanned aerial vehicles for wind turbines structure health monitoring. *Structural Health Monitoring* **2021**, 14759217211004822.
45. Márquez, F.P.G.; Chacón, A.M.P. A review of non-destructive testing on wind turbines blades. *Renewable Energy* **2020**, *161*, 998-1010.
46. Shafiee, M.; Sørensen, J.D. Maintenance optimization and inspection planning of wind energy assets: Models, methods and strategies. *Reliability Engineering & System Safety* **2019**, *192*, 105993.
47. EÓLICAS DE FUERTEVENTURA, A.I.E. PLIEGO DE PRESCRIPCIONES TÉCNICAS PARA LA PRESTACIÓN DE LOS SERVICIOS DE OPERACIÓN Y MANTENIMIENTO PREVENTIVO, PREDICTIVO Y CORRECTIVO DEL PARQUE EÓLICO DE CAÑADA DE LA

48. Zhang, Z.; Kusiak, A.; Song, Z. Scheduling electric power production at a wind farm. *European Journal of Operational Research* **2013**, *224*, 227-238.
49. Lei, M.; Shiyan, L.; Chuanwen, J.; Hongling, L.; Yan, Z. A review on the forecasting of wind speed and generated power. *Renewable and Sustainable Energy Reviews* **2009**, *13*, 915-920.
50. Weron, R. Electricity price forecasting: A review of the state-of-the-art with a look into the future. *International journal of forecasting* **2014**, *30*, 1030-1081.
51. Dai, L.; Stålhane, M.; Utne, I.B. Routing and scheduling of maintenance fleet for offshore wind farms. *Wind Engineering* **2015**, *39*, 15-30.
52. Irawan, C.A.; Ouelhadj, D.; Jones, D.; Stålhane, M.; Sperstad, I.B. Optimisation of maintenance routing and scheduling for offshore wind farms. *European Journal of Operational Research* **2017**, *256*, 76-89.
53. Christiansen, M.; Fagerholt, K.; Nygreen, B.; Ronen, D. Ship routing and scheduling in the new millennium. *European Journal of Operational Research* **2013**, *228*, 467-483.
54. Carroll, J.; McDonald, A.; McMillan, D. Failure rate, repair time and unscheduled O&M cost analysis of offshore wind turbines. *Wind Energy* **2016**, *19*, 1107-1119.
55. Martin, R.; Lazakis, I.; Barbouchi, S.; Johanning, L. Sensitivity analysis of offshore wind farm operation and maintenance cost and availability. *Renewable Energy* **2016**, *85*, 1226-1236.
56. Glassdoor. Offshore Maintenance Technician Salaries. Access date: 13/03/2018. Available online: https://www.glassdoor.com/Salaries/offshore-maintenance-technician-salary-SRCH_KO0,31.htm (accessed on
57. Puertos del estado. Punto SIMAR(1070088). Access Date: 20/03/2018. Available online: http://www.puertos.es/es-es/oceanografia/Paginas/portus.aspx (accessed on
58. Eurek, K.; Sullivan, P.; Gleason, M.; Hettinger, D.; Heimiller, D.; Lopez, A. An improved global wind resource estimate for integrated assessment models. *Energy Economics* **2017**, *64*, 552-567.
59. Taylor, J.W.; Jeon, J. Probabilistic forecasting of wave height for offshore wind turbine maintenance. *European Journal of Operational Research* **2018**, *267*, 877-890.
60. Taylor, J.W.; Jeon, J. Probabilistic forecasting of wave height for offshore wind turbine maintenance. *European Journal of Operational Research* **2017**.
61. Besnard, F.; Patriksson, M.; Stromberg, A.; Wojciechowski, A.; Fischer, K.; Bertling, L. A stochastic model for opportunistic maintenance planning of offshore wind farms. In Proceedings of the PowerTech, 2011 IEEE Trondheim, 2011; pp. 1-8.
62. Lehna, M.; Scheller, F.; Herwartz, H. Forecasting day-ahead electricity prices: A comparison of time series and neural network models taking external regressors into account. *Energy Economics* **2022**, *106*, 105742.
63. Mahler, V.; Girard, R.; Kariniotakis, G. Data-driven structural modeling of electricity price dynamics. *Energy Economics* **2022**, 105811.
64. España, R.E.d. Available online: https://www.esios.ree.es/es/curvas-de-ofertas#29?date=2018-02-16T12%3A00%20%2B0100. Access Date: 21/02/2018 (accessed on
65. Dahl, C.M.; Effraimidis, G.; Pedersen, M.H. Nonparametric wind power forecasting under fixed and random censoring. *Energy Economics* **2019**, *84*, 104520.
66. Deshmukh, M.K.; Deshmukh, S.S. Modeling of hybrid renewable energy systems. *Renewable and sustainable energy reviews* **2008**, *12*, 235-249.
67. Siemens. Wind turbine SG 8.0-167 DD. Technical Specifications. . Available online: https://www.siemens.com/content/dam/webassetpool/mam/tag-siemens-com/smdb/wind-power-and-renewables/offshore-wind-power/documents/brochures/sg-data-sheet-offshore-wind-turbine-sg-8-0-167-en.pdf (accessed on 28/05/2018).
68. García Márquez, F.P.; Peinado Gonzalo, A. A Comprehensive Review of Artificial Intelligence and Wind Energy. *Archives of Computational Methods in Engineering* **2021**, 1-24.
69. Zolpakar, N.A.; Lodhi, S.S.; Pathak, S.; Sharma, M.A. Application of multi-objective genetic algorithm (MOGA) optimization in machining processes. In *Optimization of Manufacturing Processes*; Springer: 2020; pp. 185-199.